\documentclass[a4paper,12pt]{article}
\usepackage{amsmath}
\usepackage{pstricks}
\usepackage{pst-node}
\usepackage[ansinew]{inputenc}
\usepackage{amssymb,amsmath}
\usepackage{amsfonts}
\usepackage{epsfig}
\usepackage[english]{babel}
\usepackage{graphics}

\def\p{\partial}

\def\a{\alpha}
\def\b{\beta}
\def\g{\gamma}

\def\o{\omega}
\def\e{\varepsilon}
\def\wt{\widetilde}

\font\Sets=msbm10

\def\Integer {\hbox{\Sets Z}}

\def\be{\begin{equation}}       \def\ba{\begin{array}}

\def\ee{\end{equation}}         \def\ea{\end{array}}

\def\bea {\begin{eqnarray}}      \def\eea {\end{eqnarray}}

\def\bean{\begin{eqnarray*}}    \def\eean{\end{eqnarray*}}

\def\<{\langle} \def\({\left(}  \def\>{\rangle} \def\){\right)}

\newtheorem{exi}{Example}

\newcommand{\bc}[2]{\genfrac{(}{)}{0pt}{}{#1}{#2}}

\author{Elena Kartashova, Guenther Mayrhofer\\
 RISC, J.Kepler University, Linz
4040, Austria\\
E-mails: [lena, guenther.mayrhofer]@risc.uni-linz.ac.at}

\title{Cluster formation in mesoscopic systems}

\begin{document}
\maketitle

\begin{abstract}
Graph-theoretical approach is used to study cluster formation in
mesocsopic systems. Appearance of these clusters are due to discrete
resonances which are presented in the form of a multigraph with
labeled edges. This presentation allows to construct all
non-isomorphic clusters in a finite spectral domain and generate
corresponding dynamical systems automatically. Results of
MATHEMATICA implementation are given and two possible mechanisms of
cluster destroying are discussed.


\end{abstract}


\section{Introduction}

 Mesoscopic regimes are at the frontier between classical (single
waves/particles) and  statistical (infinite number of
waves/particles) description of  physical systems. Mesoscopic
 systems is very popular topic in
various areas of modern physics and can be met in wave turbulent
theory, condensed matter (quantum dots), sociology (opinion
formation), medicine (dynamics of cardiovascular system), etc. For
instance, statistical wave turbulence theory is based on
Kolmogorov's suggestion on spatial evenness of turbulence and does
not describe observed organized structures extending over many
scales like boulders in a waterfall. Also many laboratory
experiments stay unexplained in terms of statistical approach as in
\cite{DLN06} where the experimental results have been presented for
water turbulence excited by piston-like programmed wave-makers in
water flume with dimensions 6 x 12 x 1.5 meters. The main goal of
this experiments was to establish a power-law scaling for the energy
spectrum, $\ E \sim \o^{-\nu} \ $, with some fixed $\ \nu \ $ coming
from statistical considerations and $\ \o \ $ being wave dispersion
function. It turned out that discrete effects are major and
statistical predictions are never achieved: with increasing wave
intensity the nonlinearity becomes strong before the system loses
sensitivity to the discreteness of spectral space.

Our next example is taken from a quite different area of research -
sociology. Finite size effects in the dynamics of opinion formation
have been profoundly studied  quite recently in \cite{opinion} with
essentially the same conclusions made. Namely, some changes of a
system can  be observed only when "finite number of agents in the
model takes a finite value" and thermodynamic limit does not
describe behavior of these systems. It was shown that resonance by
which a finite-size system is optimally amplified by a weak forcing
signal (identified as an advertising agent) is determined by {\it
the size} of mesoscopic system and the largest peak of the spectral
density is observed at the driving frequency.

Another interesting example of a mesoscopic system can be found in
\cite{med1} where  the flow of blood through the system of closed
tubes — the blood vessels — is described by wave equations. A
 model of the cardiovascular system as a system of coupled
oscillators is proposed and  conditions of their resonance are
studied.

Now the question is: what have boulders in waterfalls, advertisement
and cardiovascular system in common? From mathematical point of view
the answer is very simple: dynamics of all these systems can be
interpreted as {\it discrete} resonances.
 Notice that resonance conditions have the same
 general form
for wave and quantum systems (see, for instance, \cite{spohn} for
4-photon processes); and have to be studied in integers. In this
paper we have chosen a wave turbulent system as our main example and
therefore use wave terminology.

From now on we regard resonance conditions of the form
 \be \label{res} \o_1 \pm \o_2 \pm ... \pm \o_s=0, \quad
\vec{k}_1 \pm \vec{k}_2 \pm .... \pm \vec{k}_s=0 \ee where $ \
\o_i=\o(\vec{k}_i), \ \ s< \infty,\ $ with $\ \vec{k}\ $ and
$\o_i=\o(\vec{k}_i)$ being correspondingly wave vector and
dispersion function. Specific features of these systems described by
Fourier harmonics with integer mode numbers were first studied in
\cite{K94} (we call them further discrete wave systems, DWS). It is
well-known that fully statistical description of a wave turbulent
system yields wave kinetic equation \cite{ZLF} analogous to kinetic
equation known in quantum mechanics. A counter part to kinetic
equation in DWS is a set of
 few {\it independent} dynamical systems of ODEs on the amplitudes of
 interacting waves. The theory presented in \cite{K94} was based on
 a collection of pure existence theorems \cite{K98} and has been
 developed with the understanding that discrete effects are only
 important for small $\ |\vec{k}|\ $ of order $\sim 10$ while in larger
 spectral domains statistical regimes do occur. Numerous results of
 the last few years (\cite{KNP01}, \cite{LNP06}, \cite{T01a}, \cite{T01b},
  \cite{TY04}, \cite{T07}, \cite{ZKPD} just to mention a few of them)
  showed that the general conception - discrete effects are only important
  in  small spectral domains - should be revised because these
  effects
  are in fact observed in the systems where thousands of Fourier
  harmonics are taken into account, i.e. in a wide range of
  mesoscopic wave systems. A model of laminated wave turbulence has
  been presented in
  \cite{K06-1} which explains the appearance of coherent
  structures in arbitrary big but finite spectral domains. This model put a novel
 computational problem of solving (\ref{res}) in integers of order $10^3$ and more.
Fast generic algorithms for a big class of irrational and rational
dispersion functions have been developed in \cite{comp-all} which
can be used for a wide range of
 dispersion
 functions; they were also implemented (for cases $\ s=3\ $ and $\ s=4$) and  computation
time on a Pentium-3 is of order 5 to 15 minutes in
 computation domain $|m|,|n| \le 10^3,$ where $\vec{k}=(m,n), \ m,n\in
 \Integer.$ Straightforward computations for the same examples (without using our algorithms) takes few days
  with similar computer and computation domains of order $\ 10^2.$\\

In this paper we study the structure of the solution set of
(\ref{res}) using graph-theoretical approach and develop a special
 technique to construct all independent clusters and
  corresponding dynamical systems. We present the whole solution set
  as a  multigraph with labeled edges so that each connected
  component of this multigraph  correspond to a special dynamical
  system of ODEs on the wave amplitudes. The most important fact
  about is this construction is following: is provides
simultaneous isomorphism of multigraph components and dynamical
systems. Using algorithms \cite{comp-all} we have developed a
MATHEMATICA program package (at present only for $\ s=3 \ $ and
2{\bf D}-waves) capable to  1) construct
 all independent clusters of the solution set of (\ref{res})
 in a given computation domain; 2) draw them as a  multigraph on a plane;
  3) write out explicitly  all dynamical systems appearing in the
chosen spectral domain. Some results of our implementation are
given, possible directions for further research are briefly
discussed.

\section{Discrete 3-wave resonances}\label{sec:3WaveResonances}

As our main example, 3-wave resonances covered by barotropic
vorticity equation (BVE) has been chosen. This equation, also known
 as Obukhov-Charney or Hasegawa-Mima equation, is
important in many physical applications - from geophysics to
astrophysics to plasma physics: the equation was again and again
re-discovered by specialists in very different branches of physics.
In particularly, this equation
 describes ocean planetary waves \be \label{vorticity} \frac{\p
\triangle\psi}{\p x}+\b\frac{\p \psi}{\p x}  = - \e J(\psi,
\triangle\psi) \ee with non-flow boundary conditions in a
rectangular domain
$$
\psi=0 \ \ \mbox{for} \ \ x \in [0,L_x], \ y \in [0,L_y],
$$
where $\b$ is a constant called Rossby number and $\ 0 < \e \ll 1 \
$ is a small parameter.  A linear wave has then form \cite{KR92}
$$
A \cos(\frac{\b}{2\o}x+\o t)\sin{\frac{\pi m}{L_x}x} \sin{\frac{\pi
n}{L_y}y}, \ m,n\in \Integer
$$
and dispersion function can be written as
$$
\o=2/\b\sqrt{(\frac{\pi m}{L_x})^2 +(\frac{\pi n}{L_y})^2}.$$ After
obvious re-normalization we write out resonance conditions for
3-wave interactions as follows:
 \bea \label{3res_rect}
 \begin{cases}
\frac{1}{\sqrt{m_1^2+n_1^2}}+\frac{1}{\sqrt{m_2^2+n_2^2}}=\frac{1}{\sqrt{m_3^2+n_3^2}}\\
n_1 \pm n_2 = n_3
 \end{cases}
 \eea
This system will be our main subject to study in this paper.

 Just for completeness of presentation we
present here a simple idea underlying our algorithm for computing
integer solutions of (\ref{3res_rect}) (for more details see
\cite{comp-all}). It was noticed that (\ref{3res_rect}) has integer
solutions only if all three numbers $\ \sqrt{m_i^2+n_i^2},\ i=1,2,3,
\ $ have the same irrationality, i.e. can be presented as \be
\label{index} \sqrt{m_i^2+n_i^2}=\g_i\sqrt{q}\ \ee with some integer
$\ \g_i\ $ called {\it weight} and the same square-free $\ q\ $
called {\it index}. In this way, set of all wave vectors can be
divided into non-intersecting classes $Cl_q$ due to class index and
solutions are to be looked for in each class separately. it is
important to realize that
 this is only necessary condition for a solution to
exist and some classes can be empty.

As a first step, we compute the set of all possible indexes $\ q.$
Due to Lagrange theorem on presentation of an integer as a sum of
two squares we conclude that $\ q\ $ should not be not divisible by
any prime of the form $\ p=4u+3\ $ which reduces the full search
substantially. Special algorithms for
 representing square-free
numbers as sums of two squares are known, and one of them
\cite{basil} was used in numerical implementation of our algorithm
to compute the set of all possible numbers $\ q\g_i^2 \ $ such that
all $\ g_i \ $ satisfy weight equation (\ref{weight}). Special
number-theoretical considerations allowed us to disregard a lot of
classes from computations (about $\ 74\%$ of all classes in the
domain $\ m,n \le 1000$) as being empty.

Next we have to find integer solutions of the weight equation
 \be \label{weight}\frac{1}{\g_1}+\frac{1}{\g_2}=\frac{1}{\g_3}.\ee
  At this
step number of variables is  reduced from 6 to 3; their individual
degrees from 2 to 1, and we got rid from irrationality in
(\ref{weight}). Solutions are  looked for only for indexes found at
the previous step. At this step all solutions of the first equation
of (\ref{3res_rect}) are already found. Finally, we check linear
conditions on $\ n_i.$

In MATHEMATICA implementation, standard functions for list
operations and some number-theoretical function, like
\verb|SquareFreeQ| and \verb|SumOfSquaresRepresentations|, from the
standard package "NumberTheoryFunctions" are used (for details see
\cite{students}).

\section{Naive graph presentation}
\begin{figure}[h]
\begin{center}\vskip -0.2cm
\includegraphics[width=8cm,height=5cm]{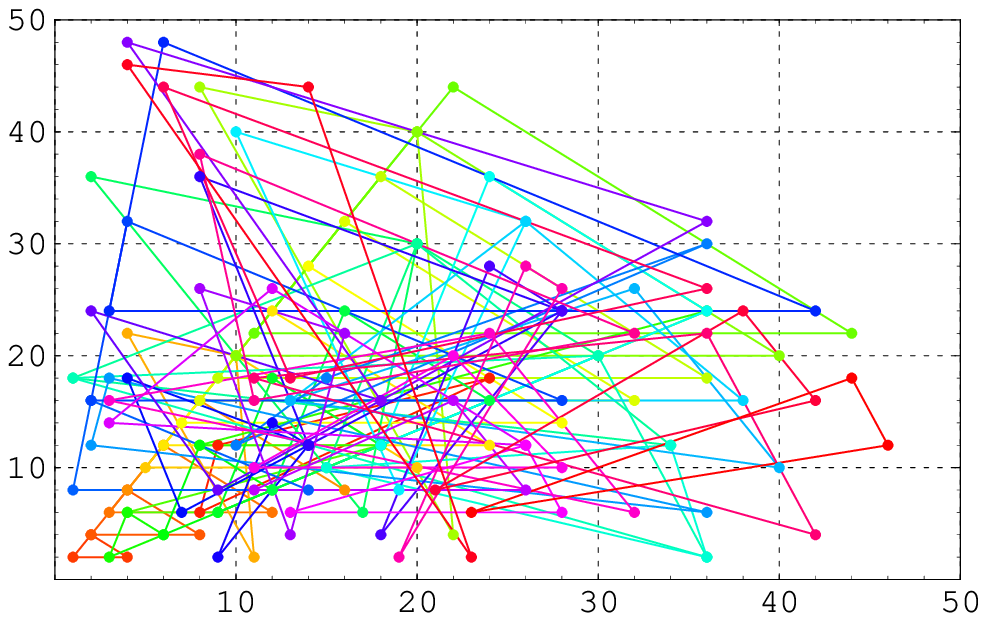}
\includegraphics[width=9cm,height=5cm]{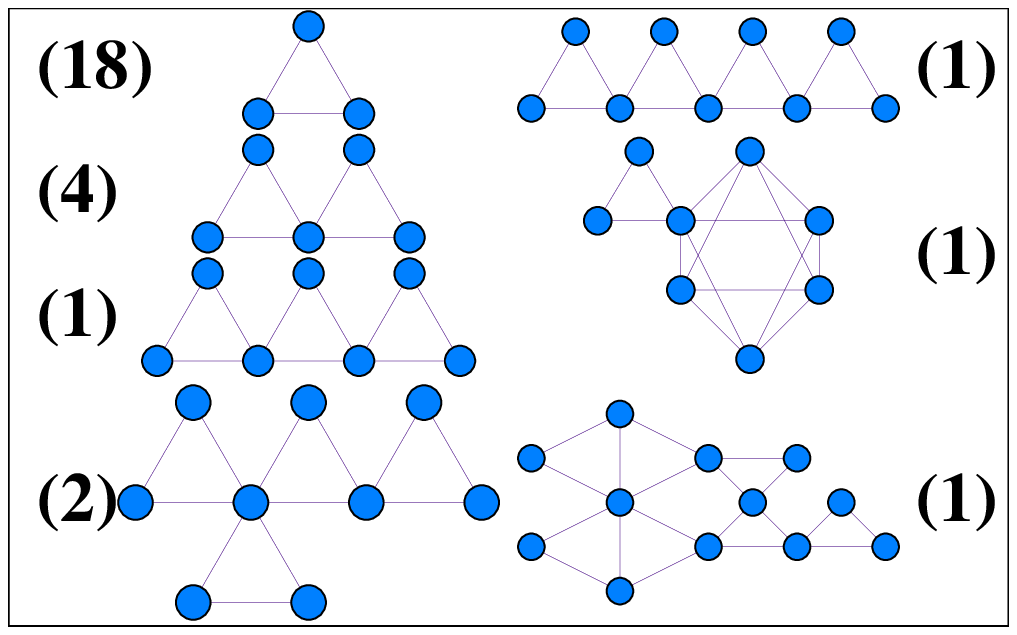}
\end{center}\vskip -0.6cm
\caption{\label{f:str_3} {\bf Upper panel:} Example of geometrical
structure, spectral domain $D=50$. {\bf Lower panel:} Example of
topological structure, spectral domain $|k_i|\le 50$. The number in
brackets shows how many times corresponding cluster appears in the
chosen spectral domain.}
\end{figure}

The graphical way to present
2D-wave resonances suggested in \cite{K98} for 3-wave interactions
is to regard each 2D-vector as a node of integer lattice in spectral
space and connect those nodes which construct one solution (triad,
quartet, etc.) We demonstrate the result in Fig.\ref{f:str_3} at the
upper panel. Obviously, geometrical structure is too nebulous to be
useful even in relatively small spectral domains. On the other hand,
topological structure shown in Fig.\ref{f:str_3} (lower panel) is
quite clear and gives us immediate information about dynamical
equations covering behavior of each wave cluster.

\begin{figure}[htb]
\begin{center}
        \begin{tabular}{cc}
            \includegraphics[height=3cm]{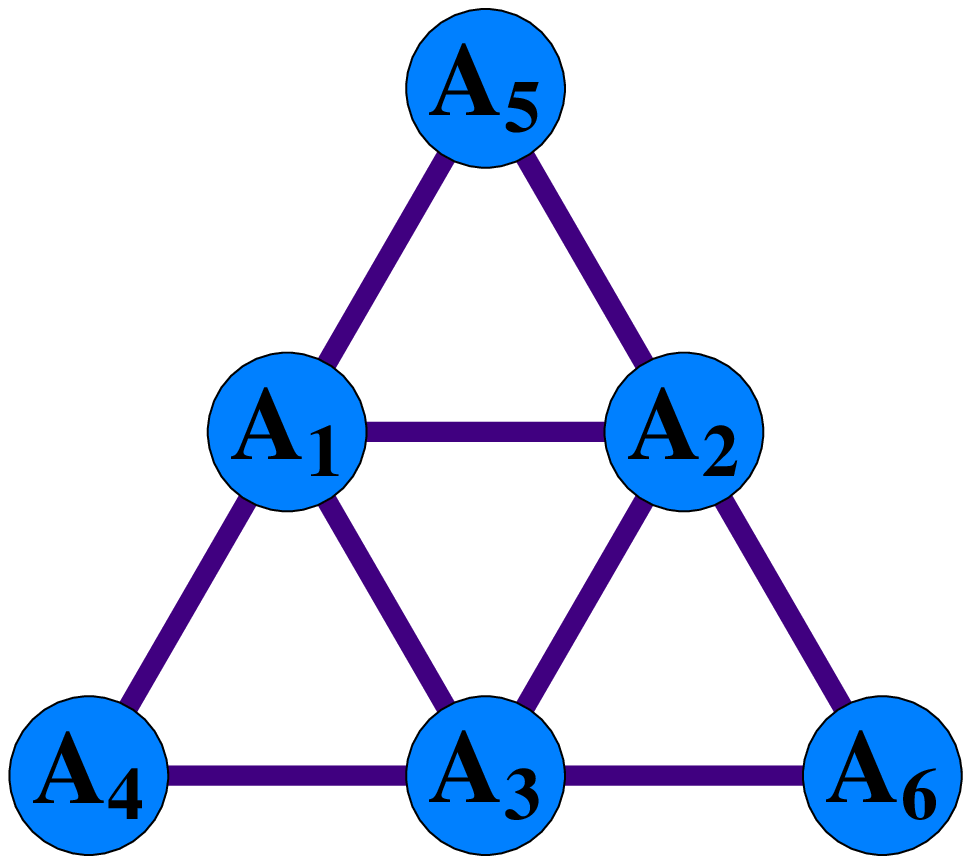} &
            \includegraphics[height=3cm]{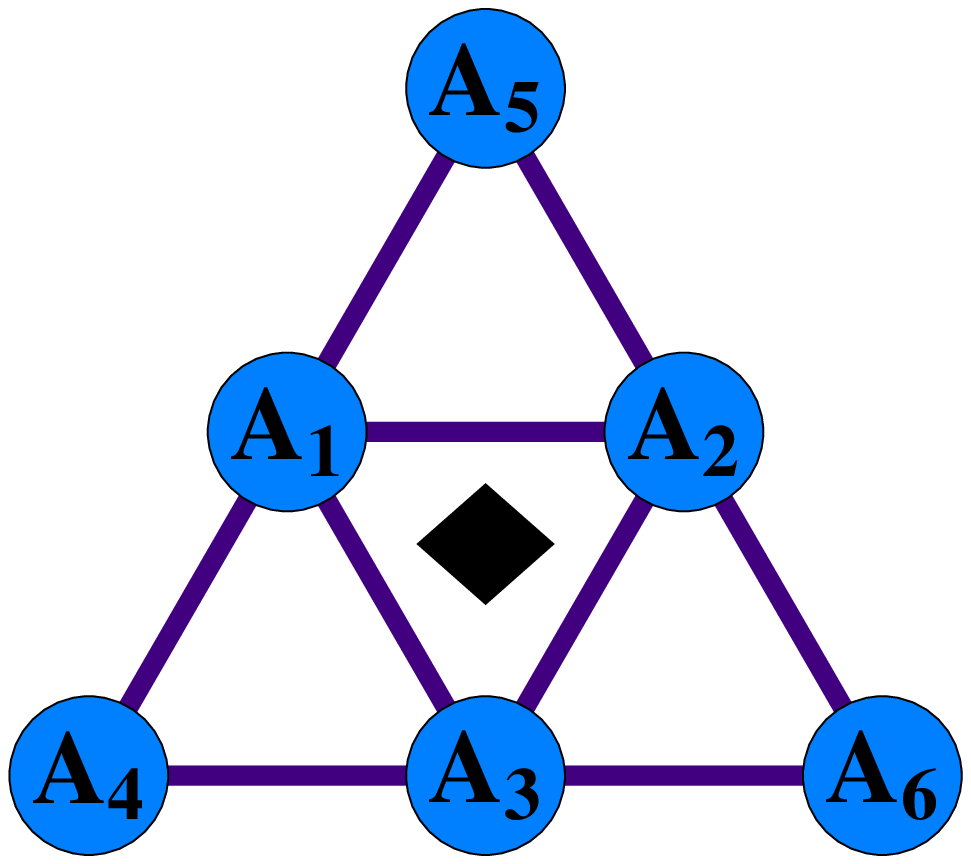} \\
        \end{tabular}
    \caption{Example of isomorphic graphs and
unisomorphic dynamical systems}
    \label{fig:HG-Example1}
\end{center}
\end{figure}

Indeed, energy transport is covered by standard dynamical system,
written for simplicity for real-valued amplitudes, \be
\label{3primary} \dot{A}_1= \a_1 A_2A_3, \
 \dot{A}_2= \a_2 A_1A_3, \
 \dot{A}_3= \a_3 A_1A_2 \
\ee in case of a "triangle" group called further {\it a primary
element}: $(A_1,A_2,A_3);$ by \be \label{3butterfly} \dot{A}_1= \a_1
A_2A_3, \
 \dot{A}_2= \a_2 A_1A_3, \
 \dot{A}_3= \frac{1}{2}(\a_3A_1A_2 +\a_4A_5A_6),\
 \dot{A}_5= \a_5 A_3A_6 , \
 \dot{A}_6= \a_6 A_3A_5\ee
 in case of "butterfly" group (two connected triangle groups):
$(A_1,A_2,A_3)(A_3,A_5,A_6)$, and so on. All isomorphic graphs
presented in Fig.\ref{f:str_3} are covered by similar dynamical
systems, only magnitudes of interaction coefficients $\a_i$ vary.
However, in general case thus defined graph structure
 does not present
dynamical system unambiguously. Consider Fig.\ref{fig:HG-Example1}
below where two objects are isomorphic {\it as graphs}. However, the
first object represents  4 connected primary elements with dynamical
system
 \be \label{dynLeft} (A_1,A_2,A_3), \ (A_1,A_2,A_5), \ (A_1,A_3,A_4), \
(A_2,A_3,A_6) \ee
 while the second - 3 connected primary elements with dynamical system
\be \label{dynRight}  (A_1,A_2,A_5), \ (A_1,A_3,A_4), \
(A_2,A_3,A_6). \ee

To discern between these two cases we set a placeholder inside
triangle not representing a resonance, we call it further empty {\it
3-cycle}. This means that to determine isomorphism of dynamical
systems we have to regard graph $G$ together with some parameter(s)
$\g$
 to identify corresponding dynamical system uniquely. We call a pair $\ (G,\g) \ $ an {\it
 i-pair} if it provides isomorphism of dynamical systems.
The set of possible parameters $\ \g\ $ (not exhaustive, of course)
is:  number of vertices, their multiplicities, number of edges,
their multiplicities, number of primary elements (non-empty
3-cycles) $\ N,$ the list of non-empty 3-cycles $\ L_c,\ $ etc. Some
preliminary study of the parameter set show that $L_c$ and $N$ is a
good first choice, providing a
 balance between informativeness and complexity of numerical
implementation.\\

Consider a structure  $(G_t, L_c)$ consisting of:
\begin{itemize}
\item{} a graph $G_t$ each edge of which belongs to at least one
3-cycle;
\item{} nonempty list $L_c$ of some 3-cycles of length $3$ of
$G_t$  such that each edge of $G_t$ belongs to some cycle(s) of
$L_c$.
\end{itemize}
Notice that $L_c$ does not contain "wrong" triangles. Notation $G_t$
has been chosen in order to point out that our graphs are, to say,
"triangle" graphs.

\paragraph{Def.1}
The number of elements in $\ L_c\ $ is called the {\it order} $\ G_t
\ $ and denoted as $\ N(G_t).\ $ 3-cycles of $\ G_t\ $ {\it not}
belonging to $\ L_c\ $ are called {\it empty cycles}. Number of
occurrences of each vertex $\ v \in G_t\ $ in $\ L_c\ $ is called
{\bf vertex multiplicity}  and denoted as
$\ \mu(v).$ Number of different vertices in $\ L_c\ $ is denoted as $\ M(G_t).$\\

Obviously,  $\ G_t$-graphs consisting of 3-cycles only, have a very
special structure. Our idea is to construct a set of all possible
graphs of this type of order $\ \le N\ $ for some given $\ N\ $
inductively, beginning with a single triangle. As a next step we can
chose all unisomorphic graphs from this set and compare
corresponding lists $\ L_c\ $ to find all different dynamical
systems.

\subsection{Triangle gluing} The possibilities of gluing a new triangle to a $\ G_t\ $-graph are
not numerous and can be classified as follows. Let the new $\ N$-th
triangle be $\ T=\{v_{1}, v_{2}, v_{3}\}.$
\begin{itemize}
\item{}Vertex gluing. In this case, $1$, $2$ or $3$ vertices of the new triangle
are identified with (glued to) vertices of some {\it distinct
triangles} of the graph, constructed at previous inductive step, $\
G_t^{(N_1)}.$
\item{}Edge gluing. In this case, $1$, $2$ or $3$ edges and corresponding vertices
of the new triangle glued to edges and vertices of some {\it
distinct adjacent triangles} of the graph. Notice that gluing of
three edges is simply filling an empty triangle of $G_t^{(N-1)}.$
\item{}Mixed gluing. In this case, one vertex $v_{N_1}$ of the new triangle
is glued to a vertices of some triangle  of the graph and the edge
$v_{N_2}v_{N_3}$ is glued to an edge of another triangle.
\end{itemize}
 The cases described
above are illustrated by figures Fig.\ref{f:gluever}-
\ref{f:gluemix}.

\newcommand{\minipagegraphics}[3]{
    \begin{minipage}{#2}
      \includegraphics[height=#3,clip]{#1}
    \end{minipage}
}

\begin{figure}[h]
\begin{center}\vskip -0.2cm
  \begin{tabular}{ccccc}
    \minipagegraphics{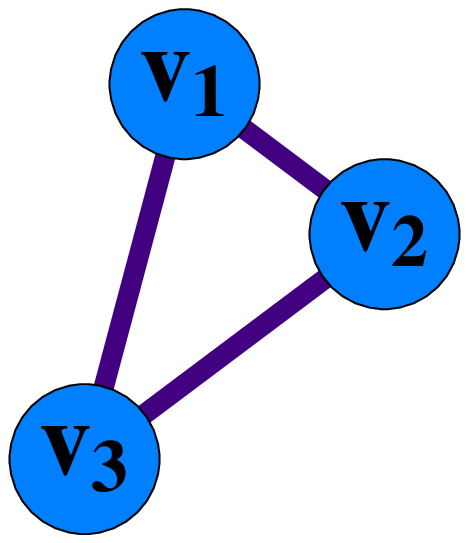}{2cm}{2.5cm}  & {\large \boldmath $+$} &
    \minipagegraphics{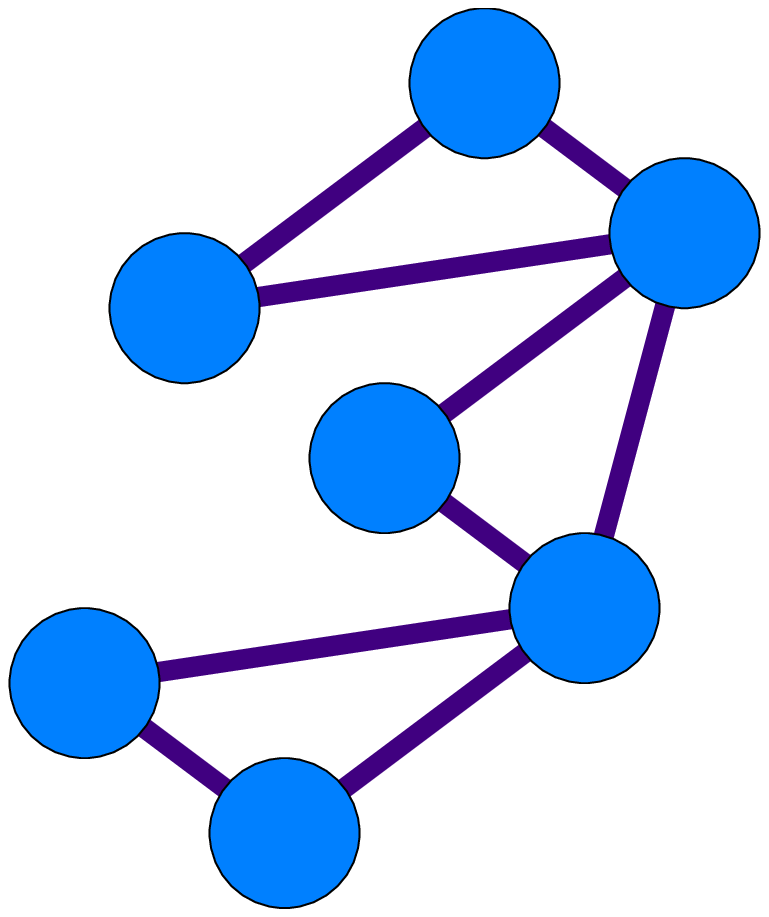}{2cm}{2.5cm,trim=1.5cm 0 0 0}  & {\large \boldmath $\Rightarrow$} &
    \minipagegraphics{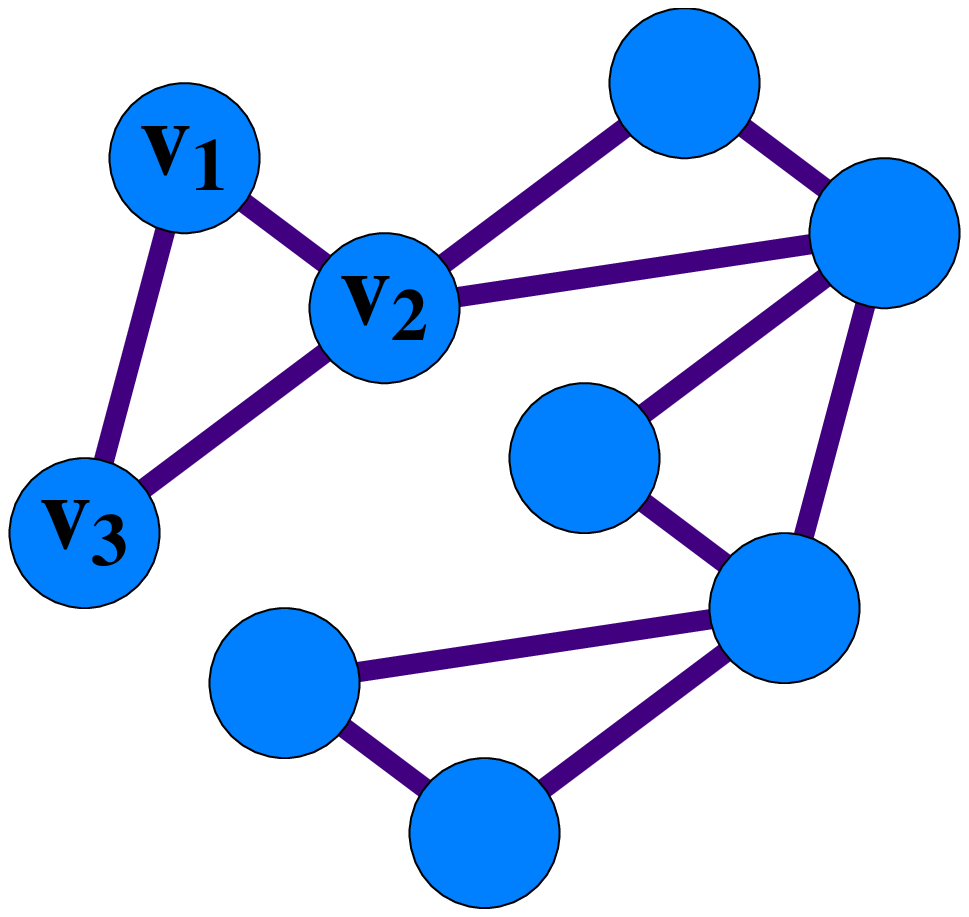}{2cm}{2.5cm}  \\
    \multicolumn{5}{c}{a) Gluing by one vertex}\\
    \minipagegraphics{VertexGluing-1.eps}{2cm}{2.5cm}  & {\large \boldmath $+$} &
    \minipagegraphics{VertexGluing-2.eps}{2cm}{2.5cm,trim=1.5cm 0 0 0}  & {\large \boldmath $\Rightarrow$} &
    \minipagegraphics{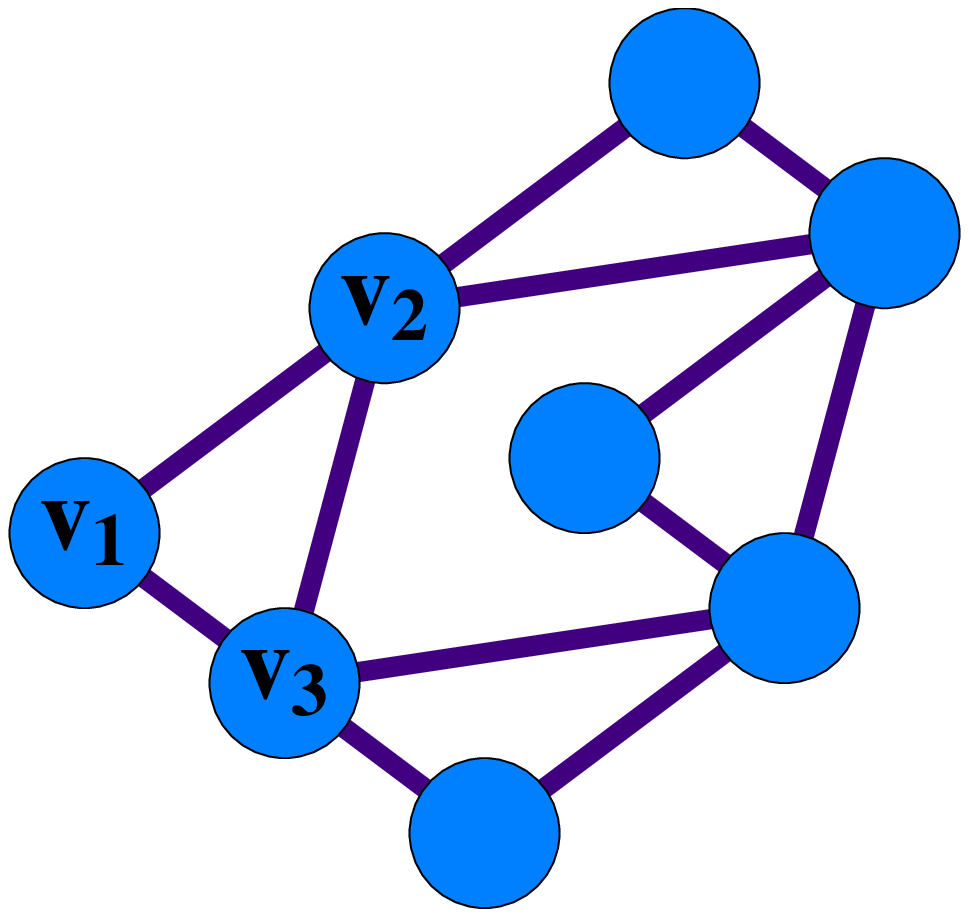}{2cm}{2.5cm}  \\
    \multicolumn{5}{c}{b) Gluing by two vertices}\\
    \minipagegraphics{VertexGluing-1.eps}{2cm}{2.5cm}  & {\large \boldmath $+$} &
    \minipagegraphics{VertexGluing-2.eps}{2cm}{2.5cm,trim=1.5cm 0 0 0}  & {\large \boldmath $\Rightarrow$} &
    \minipagegraphics{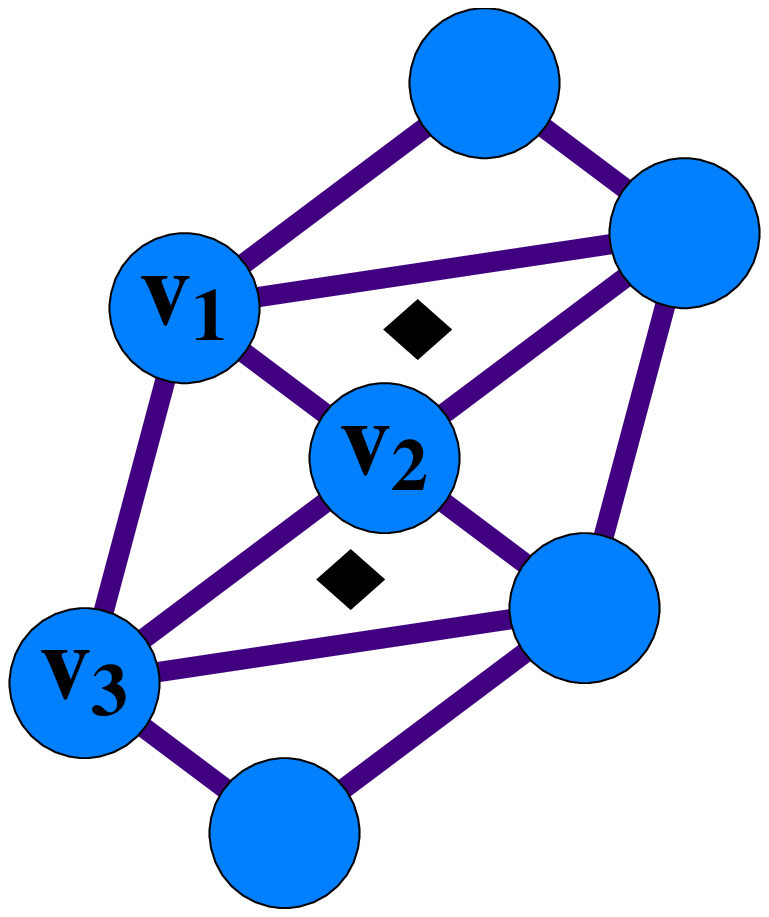}{2cm}{2.5cm}  \\
    \multicolumn{5}{c}{c) Gluing by three vertices}\\
  \end{tabular}
\end{center} \vskip -0.6cm
\caption{Vertex gluing of a new triangle to $G_t^{(N-1)}.$
\label{f:gluever}}
\end{figure}

\begin{figure}[h]
\begin{center}\vskip -0.2cm
  \begin{tabular}{ccccc}
    \minipagegraphics{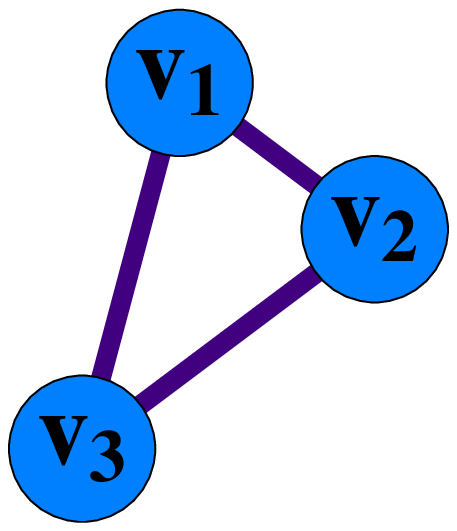}{2cm}{2.5cm}  & {\large \boldmath $+$} &
    \minipagegraphics{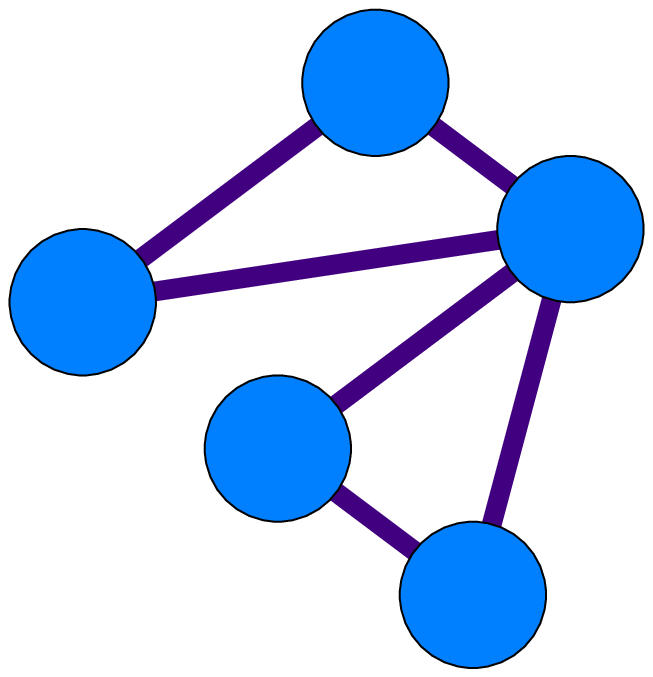}{2cm}{2.5cm,trim=1cm 0 0 0}  & {\large \boldmath $\Rightarrow$} &
    \minipagegraphics{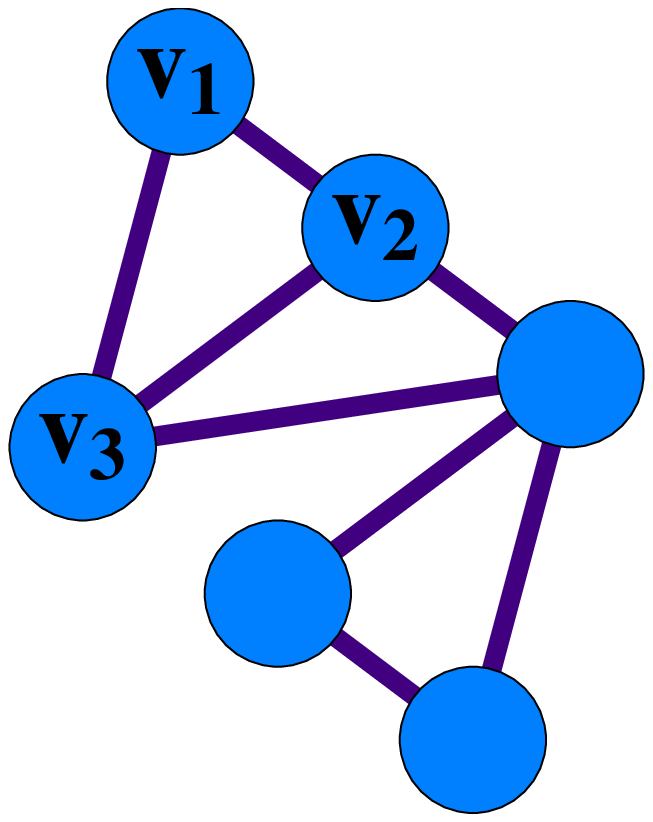}{2cm}{2.5cm}  \\
    \multicolumn{5}{c}{a) Gluing by one edge}\\
    \minipagegraphics{EdgeGluing-1.eps}{2cm}{2.5cm}  & {\large \boldmath $+$} &
    \minipagegraphics{EdgeGluing-2.eps}{2cm}{2.5cm,trim=1cm 0 0 0}  & {\large \boldmath $\Rightarrow$} &
    \minipagegraphics{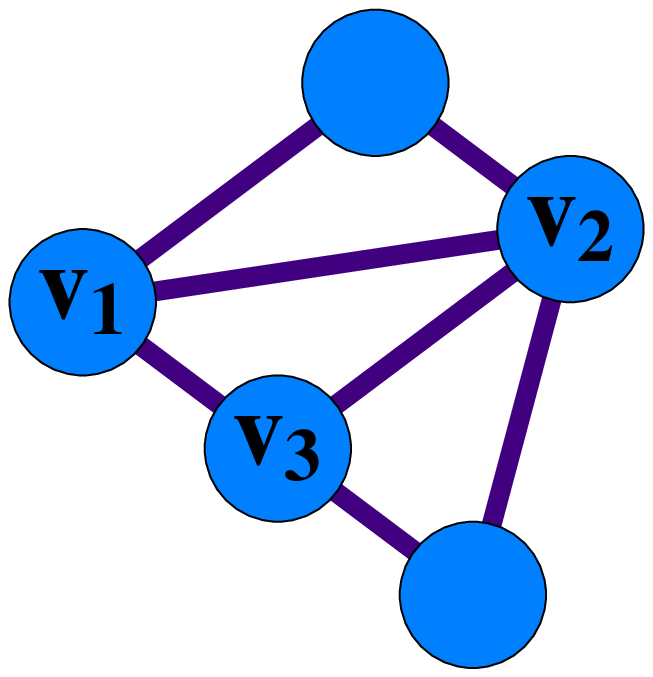}{2cm}{2.5cm}  \\
    \multicolumn{5}{c}{b) Gluing by two edges}\\
    \minipagegraphics{EdgeGluing-1.eps}{2cm}{2.5cm}  & {\large \boldmath $+$} &
    \minipagegraphics{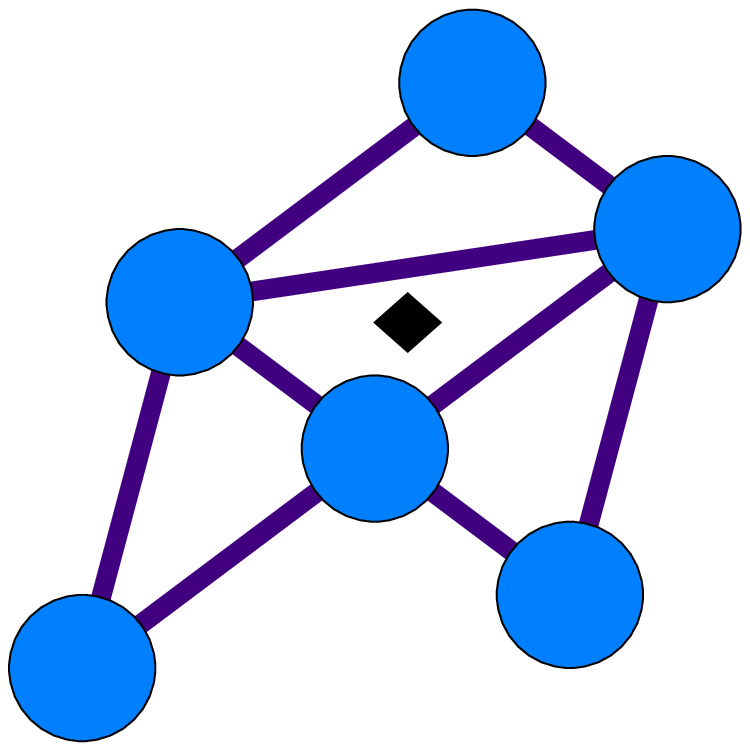}{2cm}{2.5cm,trim=1cm 0 0 0}  & {\large \boldmath $\Rightarrow$} &
    \minipagegraphics{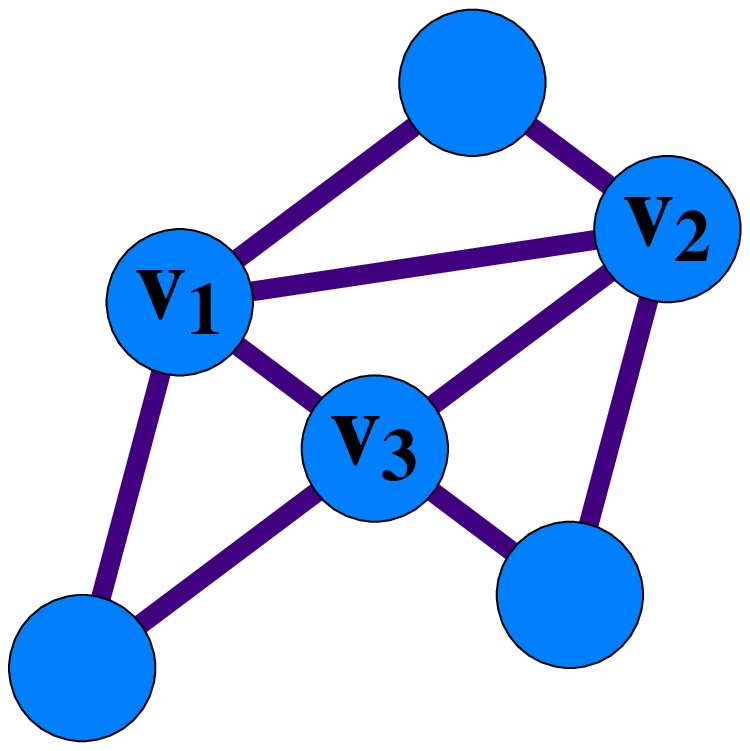}{2cm}{2.5cm}  \\
    \multicolumn{5}{c}{c) Gluing by three edges}\\
  \end{tabular}
\end{center} \vskip -0.6cm
\caption{Edge gluing of a new triangle to $G_t^{(N-1)}.$
\label{f:gluearc}}
\end{figure}

\begin{figure}[h]
\begin{center}\vskip -0.2cm
  \begin{tabular}{ccccc}
    \minipagegraphics{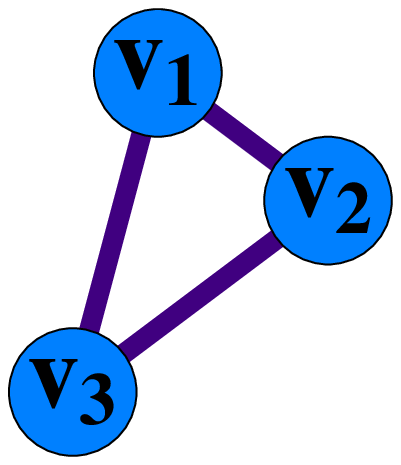}{2cm}{2.5cm}  & {\large \boldmath $+$} &
    \minipagegraphics{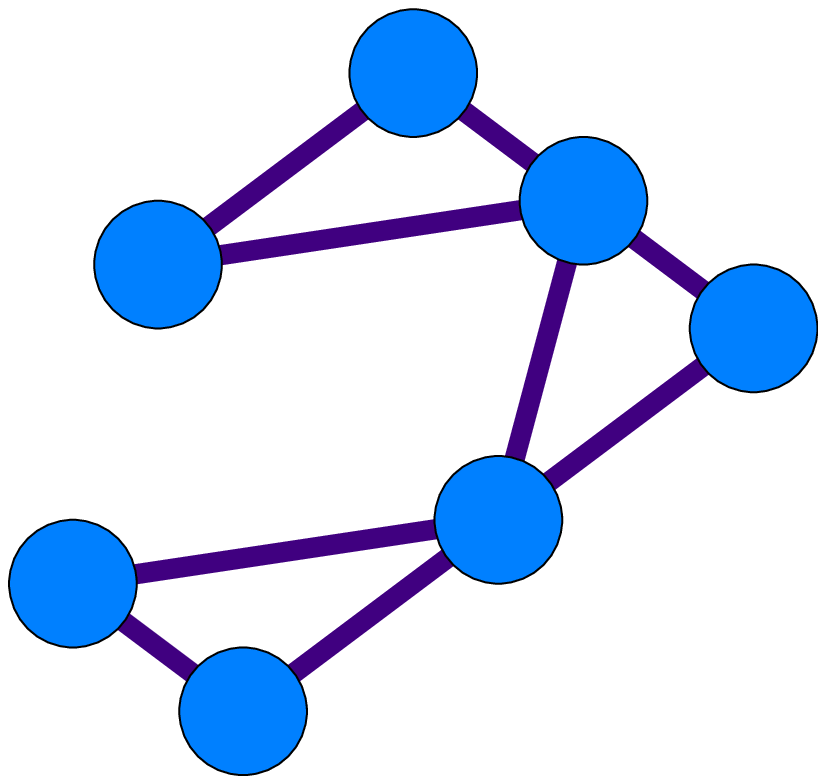}{2cm}{2.5cm,trim=1cm 0 0 0}  & {\large \boldmath $\Rightarrow$} &
    \minipagegraphics{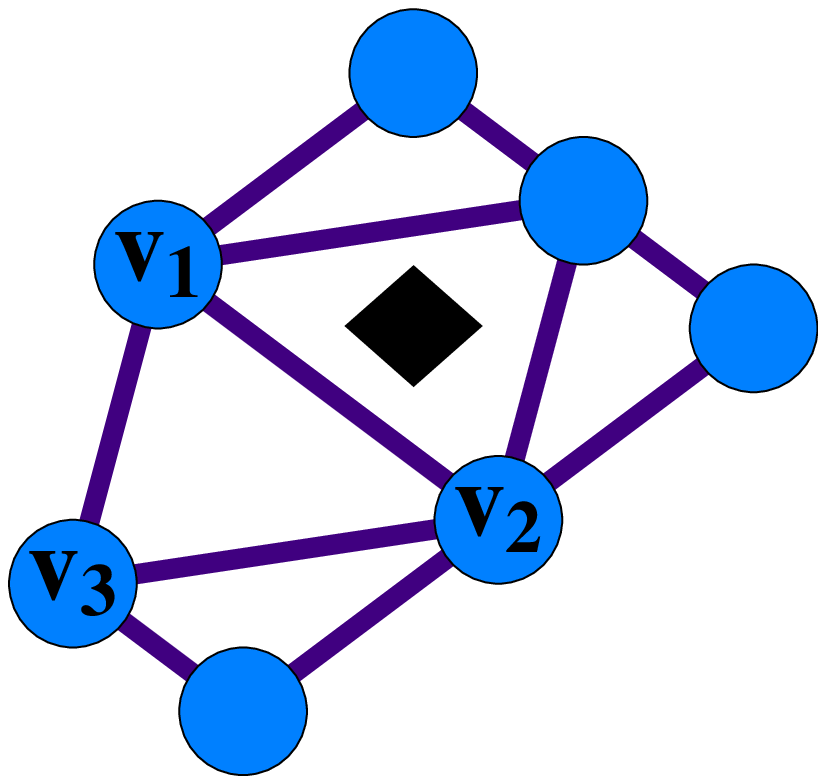}{2cm}{2.5cm}
  \end{tabular}
\end{center} \vskip -0.6cm
\caption{Mixed gluing of a new triangle to $G_t^{(N-1)}.$
\label{f:gluemix}}
\end{figure}

\begin{itemize}
\item{By vertex gluing, the $G_t^{(N-1)}$ structure is enhanced by:
   \begin{itemize}
    \item{two vertices and three edges (one vertex glued)}
    \item{one vertex and three edges (two vertices glued)}
    \item{three edges (three vertices glued)}
\end{itemize}
    }
\item{By edge gluing, the $G_t^{(N-1)}$ structure is enhanced by:
   \begin{itemize}
    \item{one vertex and two edges (one edge glued)}
    \item{one edge (two edges glued)}
    \item{the graph structure stays unchanged (three edges glued)}
\end{itemize}}
\item{By mixed gluing, the $G_t^{(N-1)}$ structure is enhanced by
two edges.}
\end{itemize}

In each case, the list $L_c^{(N-1)}$  of graph $G_t^{(N-1)}$ is
extended by the $v_{N_1}v_{N_2}v_{N_3}$ cycle of the new triangle
(or whatever these vertices will be called after the gluing).

\subsection{Some estimations}

Enhancing a given $G_t$ with $V$ vertices and $E$ edges by a
triangle, we encounter the following possibilities:
\begin{itemize}
\item{} a sole triangle not connected to the existing graph is added (1
possibility);
\item{}  vertex gluing ($V$ possibilities);
\item{}  edge gluing ($E$ possibilities);
\item{}  mixed gluing - approximately $E/V (E/V-1)$ possibilities;
\item{}  filling an empty triangle (very rare).
\end{itemize}
Therefore at each inductive step the mean number of vertices is $V
\le 1.5N$ and the number of arcs can be roughly estimated as $E \le
2N$. Therefore, the number of emerging unisomorphic graphs can be
estimated from above as some $\sim 4N$ and the overall number of
graphs at step $N$ is $\mathcal{O}(N^2)$.

\section{Hypergraph presentation}
\label{sec:HG}

To diminish computational time and complexity  we construct a {\it
hypergraph} presentation of i-pairs introduced in the previous
section. A hypergraph is a structure that consists of a set of
vertices and a multiset of edges, called hyperedges. A hyperedge is
a set of vertices, all vertices in such a set are connected. The
collection of hyperedges is a multiset because it is possible that
some hyperedges appear more times. A traditional graph is a special
case of a hypergraph, in which all edges are two-element sets and do
not appear more than once. For the representation of 3-wave
resonances we consider the triangles as "the nodes" of corresponding
hypergraph.

\paragraph{Def. 2}
A hypergraph with 3-cycles of a triangle graph $\ G_t\ $ as its {\it
vertices} and nodes $\ (m,n)\ $ of $\ G_t\ $  as its {\it edges} is
called {\it a triangle hypergraph} and is denoted as $\ HG_t.\ $ Set
of its vertices and edges is denoted as $\ V_{HG} \ $ and $\ E_{HG}
\ $
correspondingly, i.e. $\ HG_t = (V_{HG}, E_{HG}).$ \\

Notice that since a node $\ (\tilde{m},\tilde{n})\ $ of $\ G_t\ $
can belong to several 3-cycles, corresponding $\ HG_t\ $  has in
fact hyperedges instead of edges of a simple graph. A hypergraph $\
HG_t\ $     generated by $\ G_t \ $ has two properties:
\begin{itemize}
    \item{} Each vertex is part of exactly three hyperedges.
    \item{} Each pair of vertices is part of at most two hyperedges.
\end{itemize}
The first property follows from the fact, that each vertex of $\
HG_t\ $ represents a 3-cycle which consists of three different nodes
of $\ G_t.\ $  If the second property is violated then the two
associated 3-cycles of $\ G_t\ $ have three nodes in
common, hence they are identical.\\

As an illustrative example, let us write out explicitly a hypergraph
presentation of the dynamical systems (\ref{dynLeft}) and
(\ref{dynRight}) presented  in Fig.\ref{fig:HG-Example1} at the left
and right panel correspondingly: 
\be \label{leftHG}  \Big( V_{HG}= \{1,2,3,4\}, \ \ E_{HG}=\Big\{
\{2\}, \{3\}, \{4\}, \{1, 2, 3\}, \{1, 2, 4\}, \{1, 3, 4\} \Big\}
\Big)
 \ee
and
\be \label{rightHG} \Big(V_{HG}=\{1,2,3\}, \ \ E_{HG}=\Big\{ \{1\},
\{2\}, \{3\}, \{1, 2\}, \{1, 3\}, \{2, 3\} \Big \}  \Big). \ee

\subsection{Incidence matrix}
\label{sec:IncidenceMatrix} For computation purposes it is
convenient to represent a hypergraph $\ HG_t \ $ by its incidence
matrix which is constructed in the following way.

\paragraph{Def. 3} A rectangular matrix $\ \mathfrak{F} =(f_{i,j})\ $ with
$\ M(G_t) \ $ columns and $\ N(G_t)\ $ rows is called {\it incidence
matrix of} $\ G_t \ $ if 
\be \label{incMat} f_{i,j}=
\begin{cases}
1, \quad j-\mbox{th non-empty 3-cycle contains } i-\mbox{th node,}\\
0 \quad \mbox{otherwise}.
\end{cases}
\ee 
Each column of the matrix $\ \mathfrak{F} \ $ represents a triangle
in the solution set of (\ref{3res_rect}) while each row represents a
node (see Def. 1). Since we are not interested in nodes themselves
but in their relation to each other we can relabel the nodes of the
triangle with ascending integers in an arbitrary way and  use the
labels of the nodes for indexing elements in a matrix. Now
 we can construct the hyperedges of $\ HG_t \ $: if the $\ j$-th entry of a row
is equal to $1$ then we add $\ j\ $ to this hyperedge.  The vertices
of $\ HG_t \ $ are elements of $\ L_c$. The ordering of the
hyperedges is not important, because it is a multiset. However, it
is better to have a "normal form", so we sort the hyperedges by
using some ordering. Since we are interested in an implementation in
MATHEMATICA we choose the ordering used by the command \verb|Sort|.
This is an ordering, which orders lists ascending by their length,
and lists of same length lexicographical by their elements. For the
dynamical systems there is no ordering with practical advantages for
the implementation, so we let them unsorted. The incidence matrices
 of dynamical systems (\ref{dynLeft}) and (\ref{dynRight}) have form
 \be \label{leftMat}
   \left(\ba{cccc}
     1 & 1 & 1 & 0 \\
     1 & 1 & 0 & 1 \\
     1 & 0 & 1 & 1 \\
     0 & 0 & 1 & 0 \\
     0 & 1 & 0 & 0 \\
     0 & 0 & 0 & 1
   \ea\right)
 \ee
 and
 \be \label{rightMat}
   \left(\ba{ccc}
     1 & 1 & 0 \\
     1 & 0 & 1 \\
     0 & 1 & 1 \\
     0 & 1 & 0 \\
     1 & 0 & 0 \\
     0 & 0 & 1
   \ea\right)
 \ee
correspondingly. Analogously, incidence matrices of their
hypergraphs, here we use the ordering of the hyperedges described
above, 
 \be \label{leftMatHG}
   \left(\ba{cccc}
     0 & 1 & 0 & 0 \\
     0 & 0 & 1 & 0 \\
     0 & 0 & 0 & 1 \\
     1 & 1 & 1 & 0 \\
     1 & 1 & 0 & 1 \\
     1 & 0 & 1 & 1
   \ea\right)
 \ee
 and
 \be \label{rightMatHG}
   \left(\ba{ccc}
     1 & 0 & 0 \\
     0 & 1 & 0 \\
     0 & 0 & 1 \\
     1 & 1 & 0 \\
     1 & 0 & 1 \\
     0 & 1 & 1
   \ea\right)
 \ee
  are also different. The incidence matrices of the dynamical system and the corresponding
  hypergraph are not identical. The reason is the use of different orderings for the vertices.
  Matrix (\ref{leftMatHG}) is just a permuted version of matrix (\ref{leftMat}).
  Since we use a special ordering for the hyperedges, which are described by the rows of the
  incidence matrix, we obtain permuted rows.
For identifying isomorphic dynamical systems it is not necessary to
preserve an ordering, because dynamical systems with permuted
elements are still isomorphic. Hence, neither row permutations nor
column permutations destroy the isomorphism of dynamical systems. In
this example only row permutations occur, permutations of columns
are just another ordering of the elements of $\ L_c$.
 This construction can be redone and the dynamical system can be reconstructed
 out its hypergraph: by considering the columns of this matrix we know which
nodes belongs to a certain 3-cycle. \\

 Obviously, if two hypergraphs  (\ref{leftHG}) and (\ref{rightHG}) are
not isomorphic, also their incidence matrices (\ref{leftMat}) and
(\ref{rightMat}) are different. But in general
 for the
final decision it is necessary to have an algorithm to establish
isomorphism of hypergraphs. Since there are not so many general
algorithms for hypergraphs one has to find a representation where it
would be possible to use standard algorithms for graph isomorphism.
This leads us to auxiliary multigraph construction presented in the
next section.

\subsection{Multigraph construction}
\label{subsec:Multigraph}  
\begin{figure}[htb]
\begin{center}
        \begin{tabular}{cc}
            \includegraphics[width=4cm]{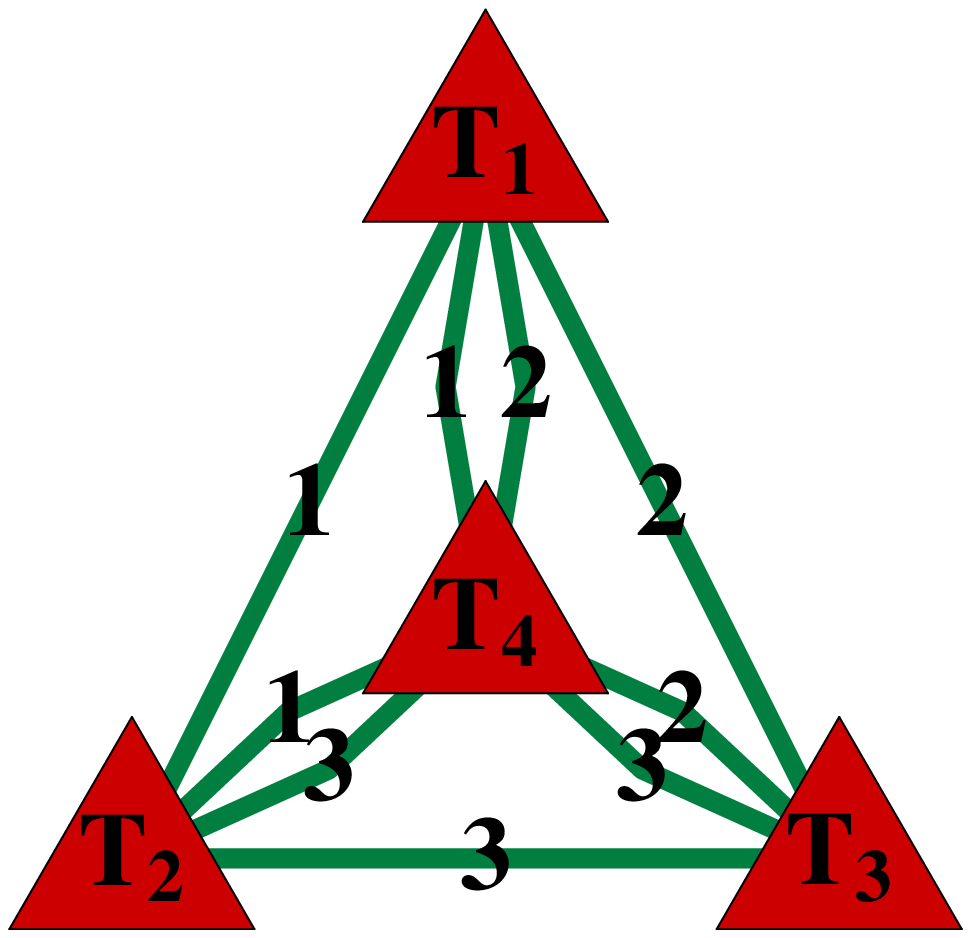} & \includegraphics[width=4cm]{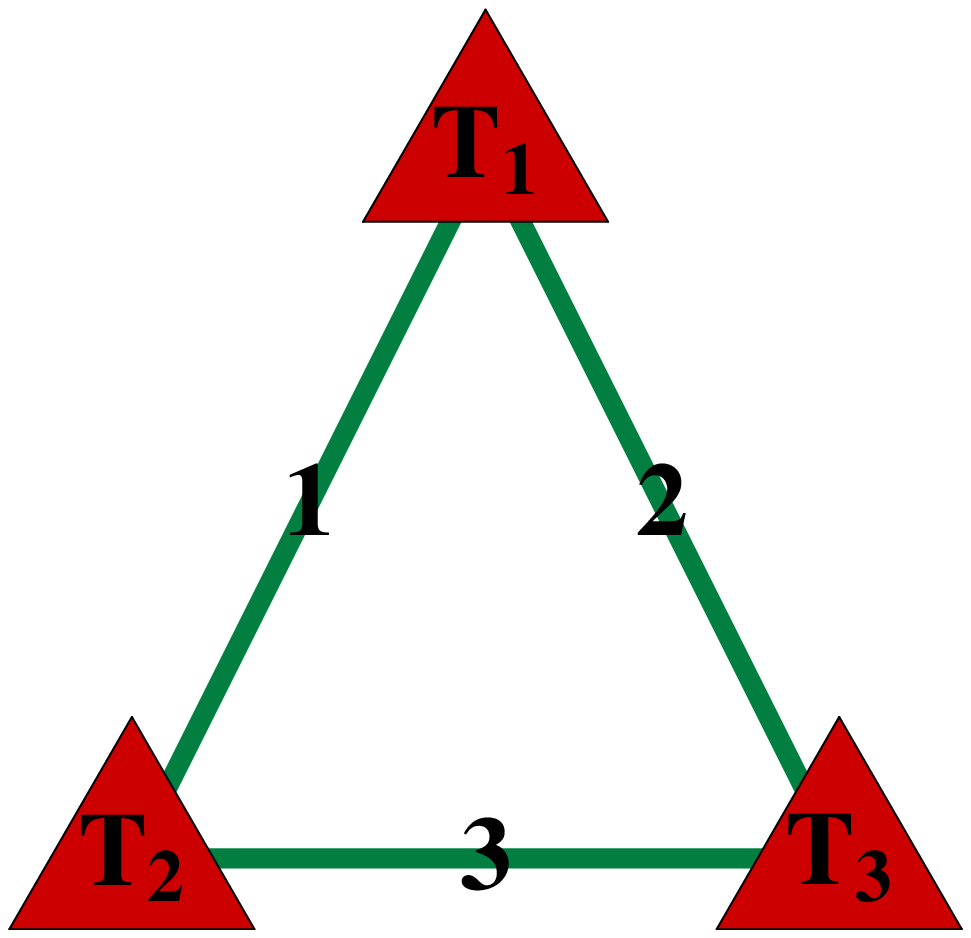}
        \end{tabular}
    \caption{Multigraph representations for dynamical systems
    (\ref{dynLeft}) and (\ref{dynRight}) correspondingly.}
    \label{fig:HG-Example2}
\end{center}
\end{figure}

A multigraph $\ MG_t \ $ is constructed
in the following way. Its vertices coincide with the vertices of $\
HG_t\ $ and each hyperedge is replaced by all two-element subsets.
To maintain the whole information we have to label the created edges
so that  edges which belong to the same hyperedge of $\ HG_t\ $ are
labeled identically. These labels allow to reconstruct $\ HG_t\ $
and $\ \mathfrak{F} \ $ which is a necessary step while generating
dynamical systems. The hyperedges which contain only one vertex can
be omitted
 because they contain no further
information about the cluster structure. Of course, some edges may
occur in $\ MG_t \ $ twice - this is the case if two 3-cycles of $\
G_t \ $ share two nodes.  Figure \ref{fig:HG-Example2} shows two
 multigraphs corresponding to the dynamical systems shown in Fig.\ref{fig:HG-Example1}. For easier distinction we use triangle
symbols for the vertices of the multigraphs, because a vertex
represents a 3-cycle of$\ G_t. $

A multigraph $\ MG_t \ $  has following properties:
\begin{itemize}
    \item{} At most two edges connect a pair of vertices.\\
{\it It follows from the fact that a pair of 3-cycles can share at
most two nodes.
    If they would share also their third node, they would be
    identical.}
    \item{} At most three different labeled edges can occur at a
    vertex.\\
    {\it A 3-cycle has three nodes therefore it can only share three different
    nodes with other 3-cycles.}
    \item{} The number of vertices is equal to the number of non-empty 3-cycles in $\ G_t.\
    $\\
    {\it By definition.}

    \item{} The total number of edges with identical labels is $\ \frac{(p-1)p}{2},\ $
    where $\ p\ $ is the number of elements in the corresponding
    hyperedge.\\
{\it Edges with identical labels belong to the same hyperedge, and
the number of two-element subsets is
       $\ \bc{p}{2}$.}
\end{itemize}

\subsection{Hypergraph {\it versus} naive graph}

Summarizing briefly the procedure described above, following has
been done:
\begin{itemize}
\item{} all integer solutions of (\ref{3res_rect}) are found;
\item{} topological presentation of the solution set as an i-pair
 $\ (G_t, \ L_c) \ $ is constructed which
presents corresponding dynamical system uniquely up to isomorphism;
\item{} i-pair $\ (G_t, \ L_c) \ $ is transformed uniquely into a hypergraph  $\ HG_t;$
\item{} for computational purposes, some auxiliary  builds  are introduced -
incidence matrix $\ \mathcal{F}(G_t)\ $ and multigraph $\ MG_t; \ $
both maintain the isomorphism of dynamical systems.
\end{itemize}

The advantages of hypergraph representation compared with a more
simple i-pair representation  given in Sec. 2 are following: 1) no
additional parameter to distinguish non isomorphic dynamical systems
are needed; 2) a standard graph isomorphism algorithm can be used to
establish the isomorphism of multigraphs; 3) the size of constructed
multigraphs is approximately one half of that for $\ G_t.\ $  Some
results of MATHEMATICA implementation of this procedure are given in
the next Section.

\section{MATHEMATICA implementation}
Details of our MATHEMATICA implementation can be found in
\cite{students} (solutions of (\ref{res}) and geometrical structure)
and in \cite{diagram} (topological structure and dynamical systems).
General computation schema is following. We implemented algorithm
sketched in Sec.\ref{sec:3WaveResonances}, computed all solutions of
(\ref{res}) and used MATHEMATICA package
"DiscreteMath`Combinatorica`" to plot triangle graph $\ G_t,\ $ and
to construct incidence matrix $\ \mathfrak{F}\ $ and multigraph $\
MG_t. \ $ To establish multigraph isomorphism we modified a standard
algorithm provided by the "DiscreteMath`Combinatorica`" package,
because it can only be used for for simple graphs and multigraphs
with unlabeled edges. Some necessary conditions of multigraphs
isomorphism are checked as a preliminary step, in order to make
computations faster. As an output, list of all resulting clusters is
given, for each of them corresponding incidence matrixes,
hypergraphs, and dynamical systems are written out, and  graphs $\
G_t\ $ and $\ MG_t\ $ are plotted. We also compute how many
isomorphic clusters of each form appear in the chosen computation
domain. Results for computation domain $\ D=50\ $ are given below.

\newcommand{\hypergraphN}[3]{
\multicolumn{2}{l}{
  \begin{minipage}[t]{\textwidth}
    #1: $ \Big( V_{HG}=\{#2\}$, $ E_{HG}=\Big\{ #3 \Big \} $ \  $ \Big ) $
  \end{minipage}
  }
}
\newcommand{\hypergraph}[3]{
\multicolumn{2}{l}{
  \begin{minipage}[t]{\textwidth}
    #1: $\Big($
    \begin{tabular}[t]{l}
    $  V_{HG}=\{#2\}, E_{HG}=\Big\{ #3 \Big \} $ \  $ \Big ) $
    \end{tabular}
  \end{minipage}
  }
}
\newcommand{\mat}[2]{
  $ \left(\begin{array}{#1} #2 \end{array}  \right) $
}
\newcommand{\dynsystem}[1]{
  \begin{minipage}{10cm}
      \begin{tabular}{l}
      #1
    \end{tabular}
  \end{minipage}
}

\begin{enumerate}

\item \begin{tabular}[t]{ll}
\hypergraph{18 systems}{1}{ \{1\},\{1\},\{1\} } \\
\mat{l}{
     1 \\
     1 \\
     1
  }  &
\dynsystem{
    $\overset{.}{A_1} = \alpha _1 A_2 A_3$ \\
    $\overset{.}{A_2} = \alpha _2 A_1 A_3$ \\
    $\overset{.}{A_3} = \alpha _3 A_1 A_2$
}
\end{tabular}
\begin{center}\vskip -0.2cm
\includegraphics[height=3.5cm]{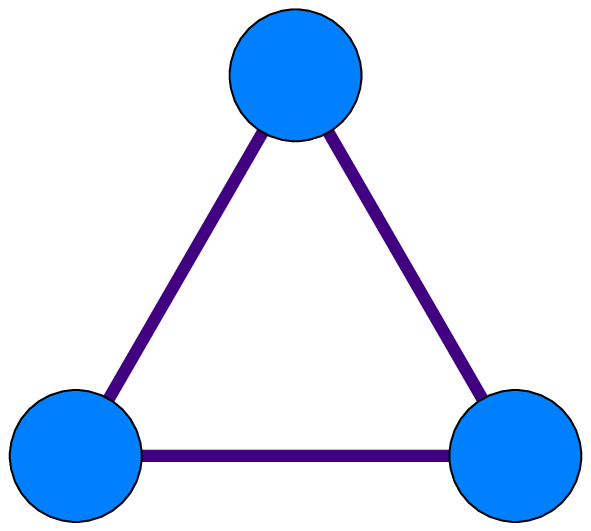}
\includegraphics[height=3.5cm]{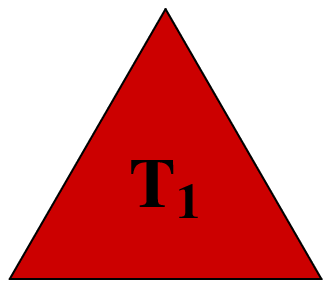}
\end{center}\vskip -0.6cm

\item \begin{tabular}[t]{ll}
\hypergraph{4 systems}{1,2}{ \{1\},\{1\}, \{2\},\{2\},\{1,2\} }
\\
\mat{ll}{
 1 & 0 \\
 1 & 0 \\
 0 & 1 \\
 0 & 1 \\
 1 & 1
  } &
\dynsystem{
    $\overset{.}{A_1} = \alpha _1 A_2 A_5$ \\
    $\overset{.}{A_2} = \alpha _2 A_1 A_5$ \\
    $\overset{.}{A_3} = \alpha _4 A_4 A_5$ \\
    $\overset{.}{A_4} = \alpha _5 A_3 A_5$ \\
    $\overset{.}{A_5} = \frac{1}{2} \left(\alpha _3 A_1 A_2 + \alpha _6 A_3 A_4 \right)$
}
\end{tabular}
\begin{center}\vskip -0.2cm
\includegraphics[width=6cm]{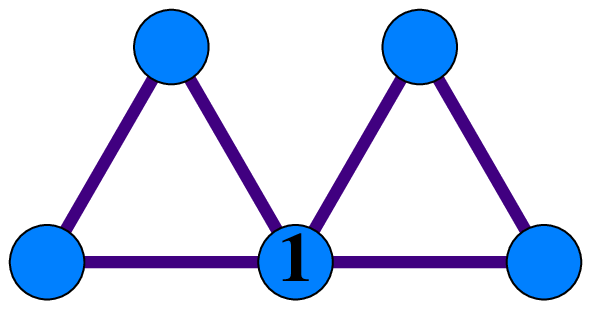}
\includegraphics[width=6cm]{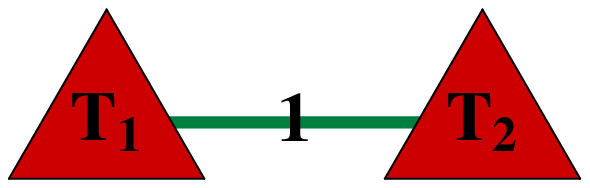}
\end{center}\vskip -0.6cm

\item \begin{tabular}[t]{ll}
\hypergraph{1 system}{1,2,3}{
\{1\},\{1\},\{2\},\{3\},\{3\},\{1,2\},\{2,3\} }
\\
\mat{lll}{
 1 & 0 & 0 \\
 1 & 0 & 0 \\
 0 & 1 & 0 \\
 0 & 0 & 1 \\
 0 & 0 & 1 \\
 1 & 1 & 0 \\
 0 & 1 & 1
  } &
\dynsystem{
    $\overset{.}{A_1} = \alpha _1 A_2 A_6 $ \\
    $\overset{.}{A_2} = \alpha _2 A_1 A_6 $ \\
    $\overset{.}{A_3} = \alpha _4 A_6 A_7 $ \\
    $\overset{.}{A_4} = \alpha _7 A_5 A_7 $ \\
    $\overset{.}{A_5} = \alpha _8 A_4 A_7 $ \\
    $\overset{.}{A_6} = \frac{1}{2} \left(\alpha _3 A_1 A_2+\alpha _5 A_3 A_7\right) $ \\
    $\overset{.}{A_7} = \frac{1}{2} \left(\alpha _9 A_4 A_5+\alpha _6 A_3 A_6\right) $
}
\end{tabular}
\begin{center}\vskip -0.2cm
\includegraphics[width=6cm]{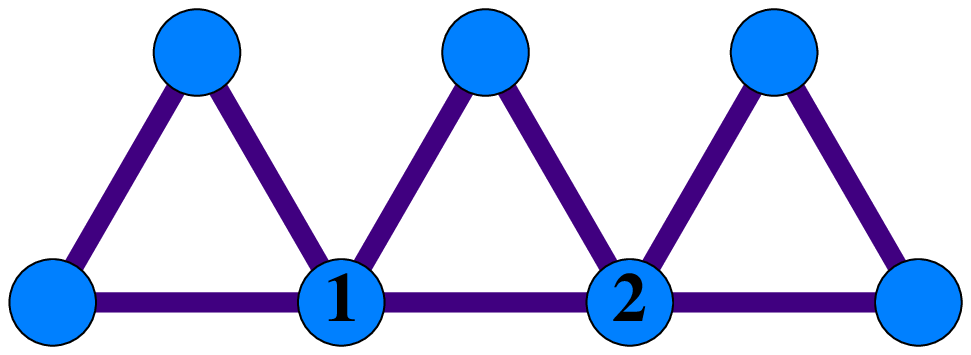}
\includegraphics[width=6cm]{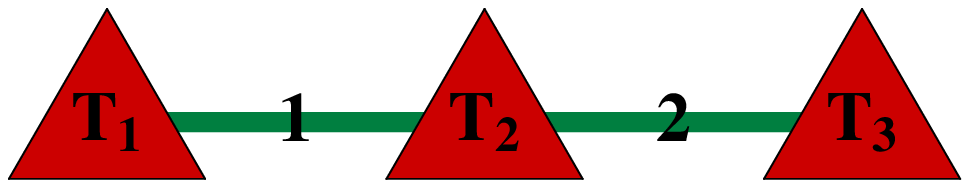}
\end{center}\vskip -0.6cm

\item \begin{tabular}[t]{ll}
\hypergraph{2 systems}{1,2,3,4}{
\{1\},\{1\},\{2\},\{2\},\{3\},\{4\},\{4\},\{3,4\},\{1,2,3\} }
\\
\mat{llll}{
 1 & 0 & 0 & 0 \\
 1 & 0 & 0 & 0 \\
 0 & 1 & 0 & 0 \\
 0 & 1 & 0 & 0 \\
 0 & 0 & 1 & 0 \\
 0 & 0 & 0 & 1 \\
 0 & 0 & 0 & 1 \\
 0 & 0 & 1 & 1 \\
 1 & 1 & 1 & 0
  } &
\dynsystem{
    $\overset{.}{A_1} = \alpha _1 A_2 A_9$ \\
    $\overset{.}{A_2} = \alpha _2 A_1 A_9$ \\
    $\overset{.}{A_3} = \alpha _4 A_4 A_9$ \\
    $\overset{.}{A_4} = \alpha _5 A_3 A_9$ \\
    $\overset{.}{A_5} = \alpha _7 A_8 A_9$ \\
    $\overset{.}{A_6} = \alpha _{10} A_7 A_8$ \\
    $\overset{.}{A_7} = \alpha _{11} A_6 A_8$ \\
    $\overset{.}{A_8} = \frac{1}{2} \left(\alpha _{12} A_6 A_7 + \alpha _{8} A_5 A_9\right)$ \\
    $\overset{.}{A_9} = \frac{1}{3} \left(\alpha _3    A_1 A_2 + \alpha _6 A_3 A_4 + \alpha _9 A_5 A_8\right)$
}
\end{tabular}
\begin{center}\vskip -0.2cm
\includegraphics[width=6cm]{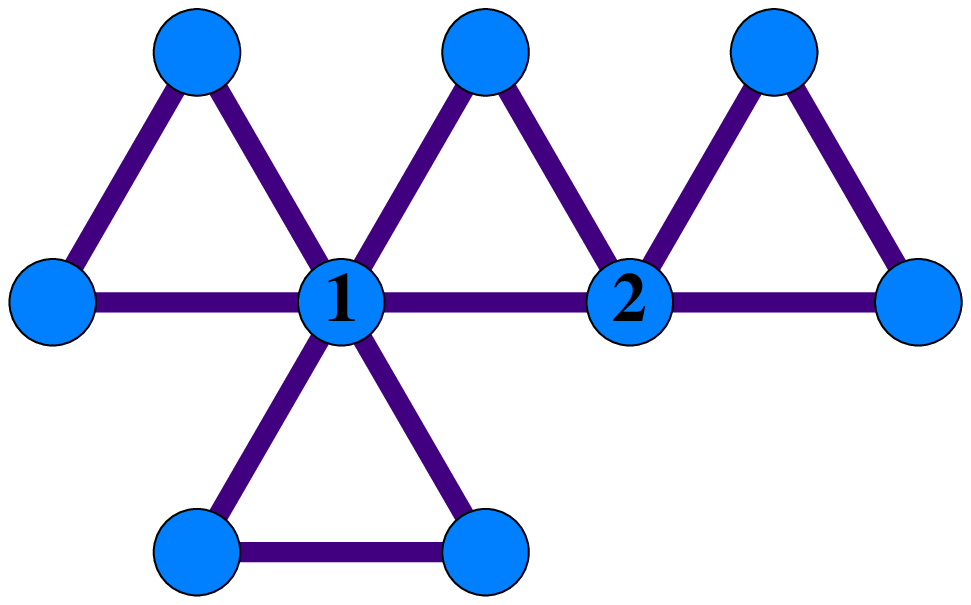}
\includegraphics[width=6cm]{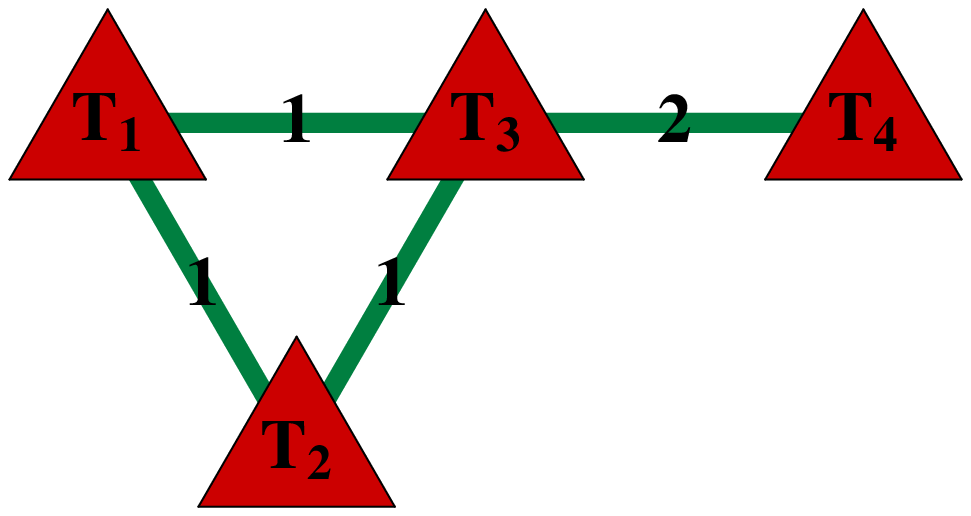}
\end{center}\vskip -0.6cm

\item \begin{tabular}[t]{ll}
\hypergraph{1 system}{1,2,3,4}{
\{1\},\{1\},\{2\},\{3\},\{4\},\{4\},\{1,2\},\{2,3\},\{3,4\} }
\\
\mat{llll}{
 1 & 0 & 0 & 0 \\
 1 & 0 & 0 & 0 \\
 0 & 1 & 0 & 0 \\
 0 & 0 & 1 & 0 \\
 0 & 0 & 0 & 1 \\
 0 & 0 & 0 & 1 \\
 1 & 1 & 0 & 0 \\
 0 & 1 & 1 & 0 \\
 0 & 0 & 1 & 1
  } &
\dynsystem{
    $\overset{.}{A_1} = \alpha _1 A_2 A_7$ \\
    $\overset{.}{A_2} = \alpha _2 A_1 A_7$ \\
    $\overset{.}{A_3} = \alpha _4 A_7 A_8$ \\
    $\overset{.}{A_4} = \alpha _7 A_8 A_9$ \\
    $\overset{.}{A_5} = \alpha _{10} A_6 A_9$ \\
    $\overset{.}{A_6} = \alpha _{11} A_5 A_9$ \\
    $\overset{.}{A_7} = \frac{1}{2} \left(\alpha _3 A_1 A_2+\alpha _5 A_3 A_8\right)$ \\
    $\overset{.}{A_8} = \frac{1}{2} \left(\alpha _6 A_3 A_7+\alpha _8 A_4 A_9\right)$ \\
    $\overset{.}{A_9} = \frac{1}{2} \left(\alpha _{12} A_5 A_6+\alpha _9 A_4 A_8\right)$
}
\end{tabular}
\begin{center}\vskip -0.2cm
\includegraphics[width=6cm]{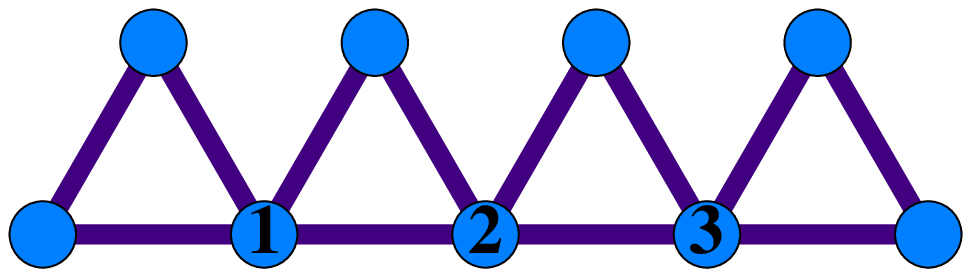}
\includegraphics[width=6cm]{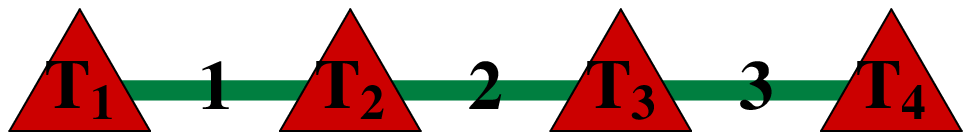}
\end{center}\vskip -0.6cm

\item \begin{tabular}[t]{ll}
\hypergraph{1 system}{1,2,3,4,5}{ \{1\},\{1\},\{2,4\},\{2,5\}, $ \\
$ \{3,4\},\{3,5\},\{4,5\},\{1,2,3\} }
\\
\mat{lllll}{
 1 & 0 & 0 & 0 & 0 \\
 1 & 0 & 0 & 0 & 0 \\
 0 & 1 & 0 & 1 & 0 \\
 0 & 1 & 0 & 0 & 1 \\
 0 & 0 & 1 & 1 & 0 \\
 0 & 0 & 1 & 0 & 1 \\
 0 & 0 & 0 & 1 & 1 \\
 1 & 1 & 1 & 0 & 0
  } &
\dynsystem{
    $\overset{.}{A_1} = \alpha _1 A_2 A_8$ \\
    $\overset{.}{A_2} = \alpha _2 A_1 A_8$ \\
    $\overset{.}{A_3} = \frac{1}{2} \left(\alpha _{10} A_5 A_7+\alpha _4 A_4 A_8\right)$ \\
    $\overset{.}{A_4} = \frac{1}{2} \left(\alpha _{13} A_6 A_7+\alpha _5 A_3 A_8\right)$ \\
    $\overset{.}{A_5} = \frac{1}{2} \left(\alpha _{11} A_3 A_7+\alpha _7 A_6 A_8\right)$ \\
    $\overset{.}{A_6} = \frac{1}{2} \left(\alpha _{14} A_4 A_7+\alpha _8 A_5 A_8\right)$ \\
    $\overset{.}{A_7} = \frac{1}{2} \left(\alpha _{12} A_3 A_5+\alpha _{15} A_4 A_6\right)$ \\
    $\overset{.}{A_8} = \frac{1}{3} \left(\alpha _3 A_1 A_2+\alpha _6 A_3 A_4+\alpha _9 A_5 A_6\right)$
}
\end{tabular}
\begin{center}\vskip -0.2cm
\includegraphics[width=6cm]{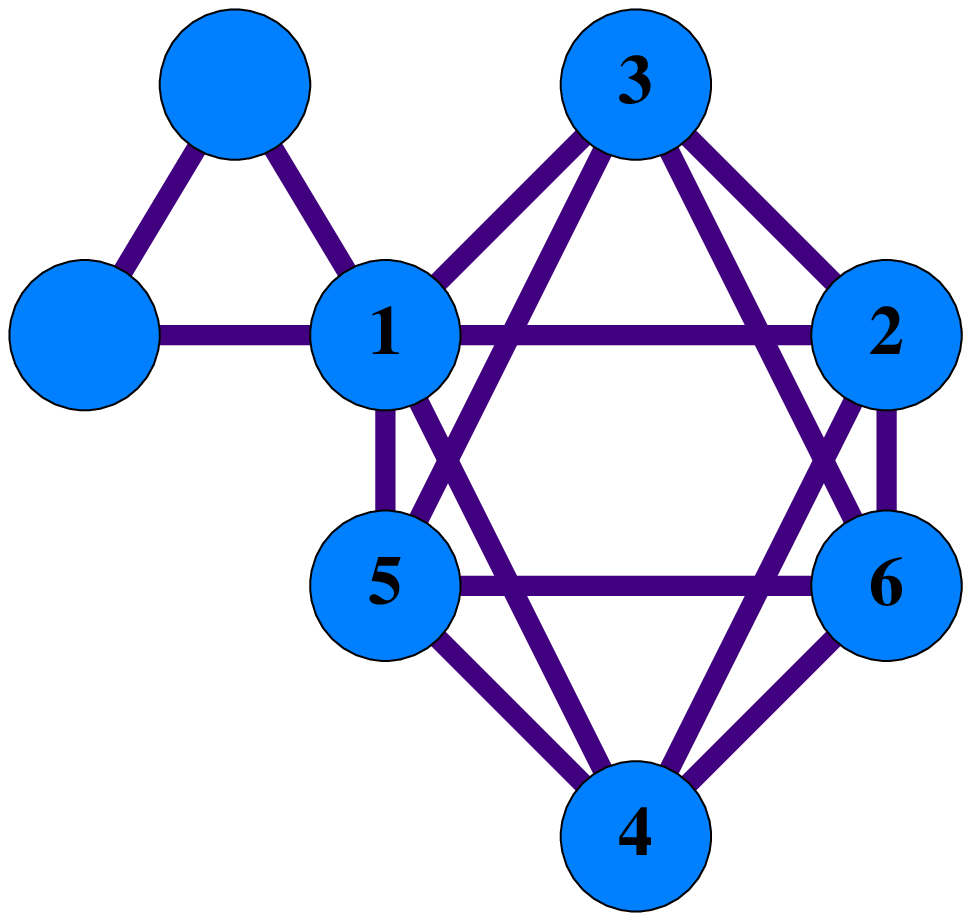}
\includegraphics[width=6cm]{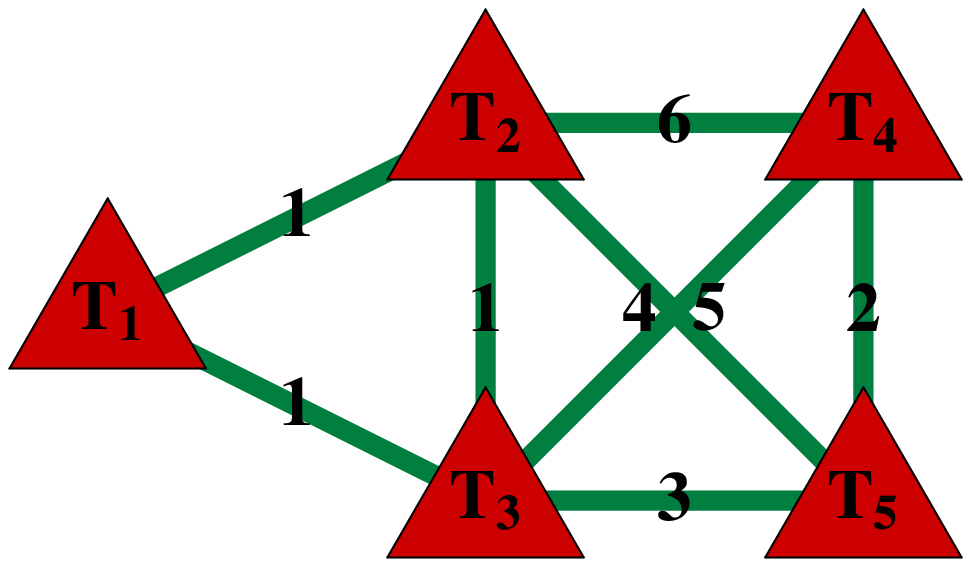}
\end{center}\vskip -0.6cm

\item \begin{tabular}[t]{ll}
\hypergraph{1 system}{1,2,3,4,5,6,7}{ \{2\},\{3\},\{4\},\{7\},\{7\},
$\\$ \{1,2\},\{1,3\},\{2,6\},\{4,5\}, \{5,6\},\{6,7\},\{1,3,4,5\} }
\\
\mat{lllllll}{
 0 & 1 & 0 & 0 & 0 & 0 & 0 \\
 0 & 0 & 1 & 0 & 0 & 0 & 0 \\
 0 & 0 & 0 & 1 & 0 & 0 & 0 \\
 0 & 0 & 0 & 0 & 0 & 0 & 1 \\
 0 & 0 & 0 & 0 & 0 & 0 & 1 \\
 1 & 1 & 0 & 0 & 0 & 0 & 0 \\
 1 & 0 & 1 & 0 & 0 & 0 & 0 \\
 0 & 1 & 0 & 0 & 0 & 1 & 0 \\
 0 & 0 & 0 & 1 & 1 & 0 & 0 \\
 0 & 0 & 0 & 0 & 1 & 1 & 0 \\
 0 & 0 & 0 & 0 & 0 & 1 & 1 \\
 1 & 0 & 1 & 1 & 1 & 0 & 0
  } &
\dynsystem{
    $\overset{.}{A_1} = \alpha _{4}  A_6 A_8$ \\
    $\overset{.}{A_2} = \alpha _{7}  A_7 A_{12}$ \\
    $\overset{.}{A_3} = \alpha _{10} A_9 A_{12}$ \\
    $\overset{.}{A_4} = \alpha _{19} A_5 A_{11}$ \\
    $\overset{.}{A_5} = \alpha _{20} A_4 A_{11}$ \\
    $\overset{.}{A_6} = \frac{1}{2} \left(\alpha _5 A_1 A_8    + \alpha _1 A_7 A_{12}\right)$ \\
    $\overset{.}{A_7} = \frac{1}{2} \left(\alpha _8 A_2 A_{12} + \alpha _2 A_6 A_{12}\right)$ \\
    $\overset{.}{A_8} = \frac{1}{2} \left(\alpha _6 A_1 A_6    + \alpha _{16} A_{10} A_{11}\right)$ \\
    $\overset{.}{A_9} = \frac{1}{2} \left(\alpha _{11} A_3 A_{12} + \alpha _{13} A_{10} A_{12}\right)$ \\
    $\overset{.}{A_{10}} = \frac{1}{2} \left(\alpha _{17} A_8 A_{11} + \alpha _{14} A_9 A_{12}\right)$ \\
    $\overset{.}{A_{11}} = \frac{1}{2} \left(\alpha _{21} A_4 A_5 + \alpha _{18} A_8 A_{10}\right)$ \\
    $\overset{.}{A_{12}} = \frac{1}{4} \left(\alpha _9 A_2 A_7 + \alpha _3 A_6 A_7 + \alpha _{12} A_3 A_9 + \alpha _{15} A_9 A_{10}\right)$
}
\end{tabular}
\begin{center}\vskip -0.2cm
\includegraphics[width=6cm]{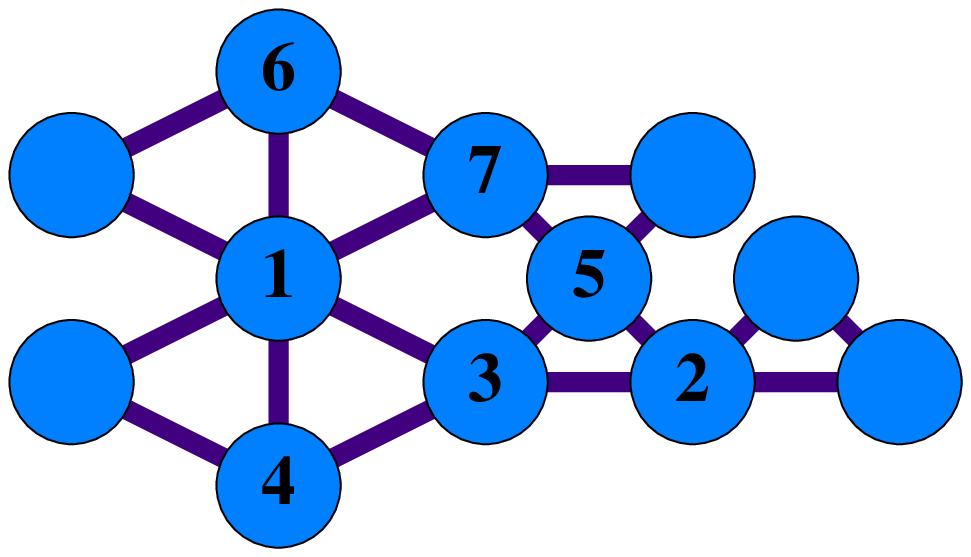}
\includegraphics[width=6cm]{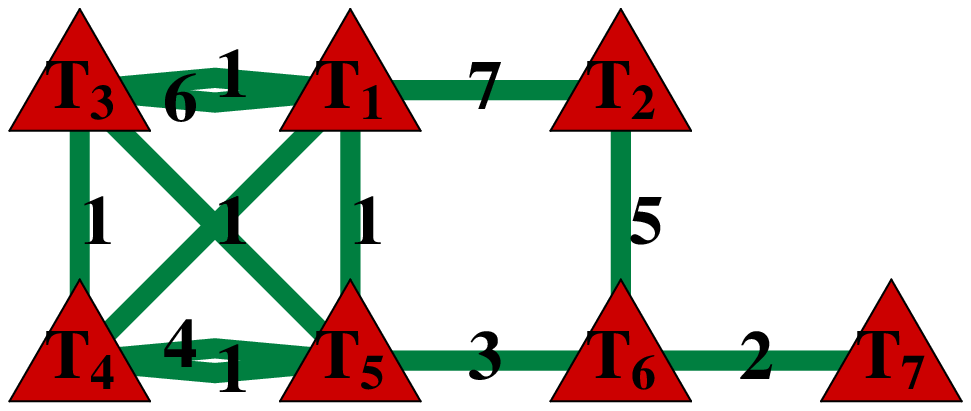}
\end{center}\vskip -0.6cm


\end{enumerate}

These results show that in spectral domain $\ D=50\ $ which contains
$\ \sim 2\cdot 10^3 \ $ Fourier harmonics we have only 7
non-isomorphic dynamical systems (clusters of waves) for further
analytical and numerical study. Some of them, for instance
(\ref{3primary}), are known to be solved explicitly in Jacobean
elliptic functions (example of explicit expressions for the case of
spherical planetary waves can be found in \cite{KL06}). Knowledge of
the explicit form of a dynamical system allows sometimes to obtain a
few conservation laws as in the case (\ref{3butterfly}) and simplify
substantially further numerical investigations of these systems. It
is important to understand that though {\it qualitative} properties
of all isomorphic clusters are the same, their {\it quantitative}
properties depend on the magnitudes of coupling coefficients $\
\alpha_i, \ $ of course. Computation of these coefficients
 is usually done
 by standard multi-scale method which is tedious but completely
 algorithmic procedure and can also be programmed in MATHEMATICA (see
 \cite{students} for its implementation).

\section{Two mechanisms to destroy clusters}

There are two mechanisms which can destroy clusters constructed
above: 1) increasing of spectral domain $\ D, \ $ and 2) taking into
account quasi-resonances, i.e. integer solutions of
\be \label{quasi}
 \o_1 \pm \o_2 \pm ... \pm \o_s=\Omega>0, \quad \vec{k}_1 \pm
\vec{k}_2 \pm .... \pm \vec{k}_s=0  \ee
with some non-zero resonance width $\ \Omega.\ $ Below we regard
briefly both of them.

\subsection{Increasing of spectral domain}
 Obviously, the structure of
clusters becomes simpler with diminishing of the domain $\ D\ $ -
some solutions (triads) disappear. On the other hand, increasing of
$\ D\ $ might lead to a substantial changes of the structure. Thus
it is important to understand how solution structure depends on the
chosen computation domain. With this aim let us re-write first
equation of (\ref{3res_rect}) in the form \be
\label{local}\frac{1}{k_1}+\frac{1}{k_2}=\frac{1}{k_3}\ee and notice
that $\ k_3 < k_1\ $ and  $\ k_3 < k_2.\ $ Introducing notations $\
k_{-}, k_{0}, k_{+}\ $ for the minimal, intermediate and maximal of
the numbers $\ k_1, k_2, k_3\ $ we conclude that $\ k_3=k_{-}.\ $
Without loss of generality one can assume that $\ k_2 \le k_1 \ $
and let $\ k_2=k_{0}\ $ and $\ k_1=k_{+}\ $ (it is a formal notation
taken for convenience of calculations, because $\ k_2\ $ and $\ k_1\
$ can also coincide). Now let us fix $\ k_{-}\ $ and re-write
(\ref{local}) as
$$
k_{+}=\frac{k_{0}k_{-}}{k_{0}-k_{-}}.
$$
Last expression achieves maximum if $\ k_0=k_- + 1$ which yields $\
k_+ \le k_-(k_- + 1) \ $ and similar considerations show that also
$\ k_0 \le k_-(k_- + 1). \ $ This means that wave interactions are
{\it local} in the following sense: lengths of wave vectors
constructing a solution of (\ref{3res_rect}) can not be too far
apart. In particularly, if we are interested in the solutions
structure in the domain, say,  $\ k_i \le D=50, \ $ it is enough to
investigate a larger domain $\  \wt{D}=50^2+50=300, \ $ in order to
establish which clusters stay unchanged and to find those which are
enhanced {\it via} solutions with wave vectors lying outside of the
initial domain $\ D=50.$

\subsection{Quasi-resonances}
It was shown in \cite{k07_grav} that for discrete quasi-resonances
to be able to start some {\it low boundary} for resonance width $\
\Omega \ $ can be written out explicitly. It interesting that for
many dispersion functions, there exist {\it global} low boundary for
most clusters which does not depend on the spectral domain under
consideration and also does not depend on the number of interacting
waves $\ s.\ $ For instance, in case of $\ \o=(m^2+n^2)^{1/4}\ $
(gravity water waves)  the use of the generalized Thue-Siegel-Roth
theorem \cite{tue} yields $\ \Omega > 1. \ $ Obviously, for
arbitrary dispersion function a {\it local} low boundary exists
which is defined   by the spectral domain  $\ T=\{(m,n): 0 < |m|,|n|
\le D < \infty\}\ $ chosen for numerical simulations. Indeed, let us
define $\  \Omega_D =\min_p \Omega_p,\ $ where 
$$
 \Omega_p =|\o(\vec k^{p}_1) \pm \o(\vec k^{p}_2) \pm ... \pm \o(\vec k^{p}_s)|,\ \
 \vec k^{p}_j = (m^{p}_j, n^{p}_j) \in T, \quad \forall j=1,2,...,s, $$
and  $$\ \o(\vec k^{p}_1) \pm \o(\vec k^{p}_2) \pm ... \pm \o(\vec
k^{p}_s) \neq 0 \quad \forall p,\ $$ and index $\ p\ $ runs over all
wave vectors in $\ T\ $, i.e. $\ p\le 4D^2\ $. So defined $\
\Omega_p\ $ obviously is a non-zero number as a minimum of finite
number of non-zero numbers and $\ \Omega_D\ $ is minimal resonance
width which allows discrete quasi-resonances to start, for chosen $\
D.$

Physically important resonance width $\ \Omega_{phys}\ $ is defined
by the accuracy of computations and precision of measurements in
numerical and laboratory experiments correspondingly.
Quasi-resonances with $\ \Omega_D\ > \Omega_{phys} $ will not
destroy the clusters.

\section{Discussion}

 In order to apply theory of discrete resonances to
a real physical problem, the profound study
 of constructed dynamical systems is needed. It is
well-known that dynamical system (\ref{3primary}) demonstrates
periodic energy exchange between the modes of a triad.  On the other
hand, dynamical systems consisting of a few connected triads have
enough degrees of freedom $\ \mathcal{N}\ $to behave chaotically.
The question of major importance therefore is to discern between two
classes of situations: 1) resonance clusters with periodic energy
exchange within each cluster, and 2) those which can be described
statistically, similar to kinetic equation approach. From this point
of view, all our theoretical results and symbolical programming can
be regarded as an introductory step for further numerical
simulations.

We are quite aware of the fact that there exists multitudinous
number of important questions to be answered in order to understand
a very complicated mutual relationship between discrete and
statistical regimes of wave system dynamics. For instance, is
corresponding statistical dynamics close to Gaussian? Is a
probability of attractor appearance in the subspace generated by
integrals of motion uniformly distributed? What is the minimal value
of $\ \mathcal{N}\ $ allowing to "forget" topological details of the
discrete, low-dimensional dynamical system and describe
corresponding dynamical system statistically? How does energy
exchange between isolated and continuous subsystems look like? What
is the role of nonlinearity in triggering energy flux toward small
scales? Is it possible to develop some analytical tools for
description of low-dimensional systems (for example, generalized
kinetic equation that accounts the finite width of frequency
resonances)? etc.

A first feeling of possible answers to some of these questions can
be obtained by computer simulations with a few well-chosen
 dynamical
systems with degrees of freedom from $\ \mathcal{N}=4\ $ to $\
10\div 20 \ $ which is in our agenda. Notice that in the case of
3-wave resonances one has
 to construct dynamical systems as it was done above and choose
 which are not enhanced by increasing of computation domain.
  Choice of initial conditions for numerical simulations
 would be another important subject to study for, as it was
 mentioned  in \cite{KL06}, even for one isolated resonant triad it
 is always possible to chose initial energy distribution among the
 modes in such a way that the period of their energy exchange will tend to
 infinity.

 In general our graph-theoretical approach can be used, with
 appropriate
 refinements, also for $\ s$-wave resonances, with $\ s\ge 4.\ $ In this last case, also
 existence of different types of resonances, spectrum anisotropy,
 etc.\cite {k07_grav} have to be taken into account in order to
 choose representative dynamical systems for numerical simulations.

 The same approach (algorithms from
\cite{comp-all}, graph construction, etc.) can also be used directly
for any mesoscopic system with resonances of a more general form
 \be
\label{res_general} p_1\o_1 \pm p_2\o_2 \pm ... \pm p_s\o_s=0, \quad
p_1\vec{k}_1 \pm p_2\vec{k}_2 \pm .... \pm p_s\vec{k}_s=0 \ee  with
integer $\ p_i\ $
\\

{\bf Acknowledgements.} Authors acknowledge the support of the
Austrian Science Foundation (FWF) under projects SFB F013/F1304 and
SFB F013/F1301.

\end{document}